\documentclass[12pt]{article}
\usepackage{amsmath}
\usepackage{cite}
\usepackage{booktabs}
\usepackage{diagbox}
\usepackage{subfig}
\usepackage[linesnumbered,ruled]{algorithm2e}
\DontPrintSemicolon 
\usepackage{graphicx}
\usepackage{amsmath}
\usepackage{lipsum}
\usepackage{multicol}
\usepackage{multirow}
\usepackage{tabularx}
\usepackage{hyperref}
\usepackage{lipsum}
\hypersetup{
	colorlinks,
	citecolor=black,
	filecolor=black,
	linkcolor=black,
	urlcolor=black
} 
\usepackage{lscape}
\makeatletter
\newcommand{\removelatexerror}{\let\@latex@error\@gobble}

\newcommand{\nosemic}{\renewcommand{\@endalgocfline}{\relax}}
\newcommand{\dosemic}{\renewcommand{\@endalgocfline}{\algocf@endline}}
\let\oldnl\nl
\newcommand{\nonl}{\renewcommand{\nl}{\let\nl\oldnl}}

\makeatother

\begin{document}
\sloppy
\definecolor{ForestGreen}{RGB}{63,142,38}

\newcommand{\fk}[1]{{\color{black}#1}}
\newcommand{\flo}[1]{{\color{black}#1}}
\newcommand{\ahmed}[1]{{\color{black}#1}}
\newcommand{\discuss}[1]{{\color{black}#1}}
\newcommand{\ali}[1]{{\color{black}#1}}
\newcommand{\aliA}[1]{{\color{black}#1}}

\newcommand{\simdag}{\mbox{SG-SD}}
\newcommand{\msg}{\mbox{SG-MSG}}
\newcommand{\smpi}{\mbox{SG-SMPI}}
\newcommand{\simgrid}{\mbox{SG}}

\title{Experimental Verification and Analysis of \\Dynamic Loop Scheduling in Scientific Applications}

\author{Ali Mohammed, Ahmed Eleliemy, Florina M. Ciorba\\
	Department of Mathematics and Computer Science\\
	University of Basel, Switzerland\\
	\and 
	Franziska Kasielke\\
	Center for Information Services and \\High Performance Computing\\
	Technische Universit{\"a}t Dresden, Germany\\
	\and
	Ioana Banicescu\\
	Department of Computer Science and Engineering\\
	Mississippi State University, USA\\
}

%
%
%
%
%
%
 \maketitle
\clearpage
\tableofcontents
\clearpage

\begin{abstract}

Scientific applications are often irregular and characterized by large \aliA{computationally-intensive} parallel loops. 
Dynamic loop scheduling (DLS) techniques can be used to improve the performance of computationally-intensive scientific applications via load balancing \aliA{of} their execution on high-performance computing~(HPC) systems. 
Identifying the most suitable choices of data distribution strategies, system sizes, and DLS techniques which \aliA{improve the performance of} a given application, requires intensive assessment and a large number of exploratory native experiments (using real applications on real systems), which may not always be feasible or practical due to associated time and costs.
In \aliA{such cases}, to avoid the execution of a large volume of exploratory native experiments, simulative experiments which are faster and less costly are more appropriate for studying the performance of applications for the purpose of optimizing it.
This motivates the question of \emph{`How realistic are \aliA{the} simulations of executions of scientific applications using DLS on HPC platforms?'}
In the present work, a methodology is devised to answer this question
\aliA{It} involves \aliA{the} experimental verification and analysis of the performance of DLS in scientific applications. 
The proposed methodology is employed for a computer vision application executing using four DLS techniques on two different HPC platforms, both via native and simulative experiments. 
Moreover, the evaluation and the analysis of the native and simulative results indicate that the accuracy of the simulative experiments is strongly influenced by the values obtained by the chosen approach used to extract the computational effort of the application~(FLOP- or time-based), the choice of application model representation into simulation (data or task parallel), and the choice of HPC subsystem models available in the simulator~(multi-core CPUs, memory hierarchy, and network topology). 
Further insights into the native performance on two HPC platforms one versus the other, 
the simulated performance using the two SimGrid interfaces, one versus the other, and
the native versus the simulated performance for each of the simulated HPC platforms, are also presented and discussed. 
\aliA{The minimum and the maximum percent errors achieved between native and simulative experiments are~$0.95\%$ and~$8.03\%$, respectively.}

%

\end{abstract}


\section{Introduction}
\label{sec:intro}

In scientific applications, loops are prevalent and represent the primary source of parallelism, most of the application execution time being spent executing loops. 
Loops of scientific applications \aliA{typically} have a large number of iterations.
During the execution of loop iterations, an  imbalanced load execution \aliA{due to problem, algorithmic or systemic characteristics,} may result in a performance degradation of the entire application.
Static and dynamic loop scheduling~(DLS) techniques play an essential role in \aliA{improving} the performance of scientific applications. 
These techniques balance the assignment and the execution of different loop iterations across the available computing units.
Hence, this balancing maximizes the application's performance. 
Throughout the present work, a loop iteration refers to a task, both terms being used interchangeably.
Identifying the best choices among available DLS techniques for a given application requires intensive assessment and a large number of native experiments\footnote{\aliA{Native experimentation is oftentimes also referred to as ``direct'' or ``bare~metal'' experimentation in the literature, and denotes experiments using real applications on real computing systems}}.
This significant amount of experiments may not always be feasible or practical due to their associated time and costs.	 
Simulation mitigates such costs and, therefore, has been shown to be more appropriate to study the performance for the purpose of optimizing it.

The performance of scientific applications is commonly studied natively on real high performance computing (HPC) platforms or using simulators, with simple and straightforward loop scheduling techniques being considered.
However, to promote the trustworthiness of the performance insights obtained via simulation, the simulated performance of an application should be studied, analyzed and compared to the native performance of the application. 
\aliA{An important source of uncertainty in the performance results obtained via simulation is the degree of trustworthiness in simulation. 
Attaining a high degree of trustworthiness eliminates this source for the present and further experiments with more complex scheduling techniques and with variable processor and network availabilities.}
The absence of such analyses motivates the question of `\emph{How realistic are \aliA{the} simulations of executions of scientific applications using DLS on HPC platforms?}'

In the present work, a methodology is proposed to experimentally verify and analyze the performance of scientific applications using DLS both, natively and via simulation in an \aliA{effort} to answer the above question.
A review \aliA{of} DLS techniques can be found in the literature~\cite{Banicescu2005}.
The proposed methodology is built upon three perspectives of comparison of the results of native and simulative experiments: native-to-native, simulative-to-simulative, and simulative-to-native.
Through the first perspective, the performance of an application executed on different parallel and distributed platforms using different DLS techniques is examined and compared.
This comparison is essential to understand the main characteristics of the application and the application sensitivity to different hardware specifications of the HPC platforms.
In the second comparison perspective, the representation of the application and the platform characteristics are used in different simulators which implement DLS techniques. 
The simulated performance of the application is compared among the different available DLS techniques.
Given that the simulators use the same application and platform characteristics, this comparison allows a better assessment of the simulator's influence on the performance.    
In the third perspective, the performance of the native experiments is compared against the one of the simulative experiments.
Based on the information obtained from the first two comparisons, this comparison is to verify and justify the level of the agreement between the results of native and simulative experiments.

The proposed methodology is applied to an application from computer vision.
This application uses the parallel spin-image algorithm (PSIA)~\cite{psia}, an enhanced version of the \mbox{well-known} \mbox{spin-image} algorithm~(SIA)~\cite{johnson1999using}.
This algorithm has various applications, such as, 3D object recognition, categorization, and face recognition.
The usage of the PSIA is considered as an example of scientific applications where the performance of a single large loop dominates the entire execution of the application.
Loop scheduling techniques have the potential to enhance the performance of the \mbox{PSIA}~\cite{Eleliemy:2017b}.
In this work, the performance of the PSIA using different loop scheduling techniques, namely static~(STATIC), \mbox{self-scheduling}~(SS)~\cite{SS}, fixed size chunking~(FSC)~\cite{FSC}, guided \mbox{self-scheduling}~(GSS)~\cite{GSS}, and factoring~(FAC)~\cite{FAC}, is analyzed, both natively on two HPC platforms (miniHPC and Taurus), and via two \mbox{SimGrid}~(\simgrid{})~\cite{SimGrid} simulation interfaces: \mbox{MetaSimGrid}~(\msg{}) and SimDag~(\simdag{}).

The present work makes the following contributions: 
(1)~Offers a method for obtaining high confidence in the results obtained both, natively and via simulation;
(2)~Provides an experimental verification and validation of the use of the different SG interfaces to represent the application loop characteristics for the purpose of developing and testing DLS techniques in simulation;
(3)~Evaluates the usefulness of using the floating point operations (FLOP) count vs. time-based measurements to represent the application characteristics in simulation; \aliA{and} 
(4)~The results of  the verification and the analysis strongly indicate that the absence of modern CPU and memory models in SG \aliA{may adversely influence} the close agreement between native and simulative experimental results, and thus, the relevance of the insights reported regarding the application performance in simulation.



The rest of this work is organized as follows. 
In Section~\ref{sec:background}, a review is presented of the selected DLS techniques, the \simgrid{}  simulation toolkit, as well as of the relevant research efforts regarding the use of simulation in performance studies of loop scheduling techniques.
The description of the selected parallel application and HPC systems is given in Section~\ref{sec:app}.
The proposed methodology for experimental verification and analysis of DLS is introduced and discussed in Section~\ref{sec:method}.
The experimental results of executing PSIA using DLS, both natively and via simulation are presented and discussed in Section~\ref{sec:predict}.
The conclusions and the potential future work are outlined in Section~\ref{sec:con}.

\section{Background and Related Work}
\label{sec:background}

This section consists of three parts. The first two parts discuss essential concepts concerning the loop scheduling techniques and the \simgrid{} simulation toolkit.
The last part includes a review of the relevant research efforts on the use of simulation in studying the performance of loop scheduling techniques.\\

\noindent\textbf{Loop scheduling.}
There are two main categories of loop scheduling techniques: static and dynamic. 
The essential difference between static and dynamic loop scheduling is the time when the scheduling decisions are taken.
Static scheduling techniques, such as block, cyclic, and block-cyclic~\cite{li1993locality}, divide \aliA{and assign} the loop iterations (\aliA{or} tasks) across the processing elements before the loop executes.
\aliA{The tasks division and assignment do not change during execution. 
In this work, block scheduling is considered and is denoted as STATIC.}

\aliA{Dynamic loop scheduling (DLS)} techniques divide and \aliA{\mbox{self-schedule}} the loop iterations during the execution of the loop. 
As a result, DLS techniques balance the execution of the loop iterations at the cost of increased overhead compared to the static techniques.
DLS techniques consider independent loop iterations (\aliA{or} independent tasks) of applications.
\aliA{For dependent tasks,}, several loop transformations, such as loop peeling, loop fission, loop fusion, and loop unrolling can be used to eliminate loop dependencies~\cite{surveyloop}.
DLS techniques can be categorized as \emph{non-adaptive} and \emph{adaptive}.  
During the application execution, the \mbox{non-adaptive} techniques calculate the chunk sizes based on certain parameters that can be obtained prior to the application execution.
The adaptive DLS techniques exploit during execution the latest information on the state of both the application and the system to predict the next sizes of the chunks of the iterations to be executed.  
In highly irregular environments, the adaptive DLS techniques balance the execution of the loop iterations \aliA{significantly} better than the \mbox{non-adaptive} techniques.
However, adaptive techniques may result in significant scheduling overhead compared to the \mbox{non-adaptive} techniques and \aliA{are,} therefore, recommended in cases characterized by highly imbalanced execution. 

This work considers the \mbox{non-adaptive} DLS techniques \aliA{while the adaptive techniques are planned as future work.}
In particular, SS~\cite{SS}, FSC~\cite{FSC}, GSS~\cite{GSS}, and FAC~\cite{FAC} are used herein. 
SS~\cite{SS} is one of the simplest \aliA{DLS} techniques.
When a processing element becomes free and available, it \aliA{retrieves} a single loop iteration from a central work queue.
\aliA{In general,} SS can achieve \aliA{a high} load balancing between all processing elements. 
However, this advantage is at the cost of  higher execution overhead compared to other DLS techniques.
FSC~\cite{FSC} \aliA{avoids the large overhead of single loop iterations being retrieved at a time by grouping iterations into chunks at each scheduling round}.  
In FSC, the chunk size is fixed and plays a critical role in determining the performance of this technique. 
FSC needs profiling to obtain certain information such as the mean of iterations' assignment overheads and the standard deviation of loop iteration execution times.
GSS~\cite{GSS} and FAC~\cite{FAC} \aliA{are improvements to SS in terms of decreased scheduling overhead at decreased load balancing.}
GSS divides the total number of loop iterations into \mbox{decreasing-sized} chunks.
\aliA{Upon a work request,} the remaining loop iterations are divided by the total number of processing elements.
FAC \aliA{improves} GSS by scheduling the loop iterations in batches of \mbox{equal-sized} chunks. 
The initial chunk size of GSS is usually larger than the size of the initial chunk using FAC.
If more \mbox{time-consuming} loop iterations are at the beginning of the loop, FAC \aliA{balances} the execution better than GSS. 
\aliA{The chunk calculation in FAC is based on probabilistic analysis to balance the load among the processes, depending on the prior knowledge of the mean and the standard deviation of the loop iterations execution times.
	Due to the fact that loop characteristics are not known apriori and typical loop characteristics that can cover many probability distributions, a practical implementation of FAC was suggested~\cite{FAC} that assigns half of the remaining work in a batch. 
	This work considers this practical implementation.} 
Compared to STATIC and SS, GSS and FAC provide better \mbox{trade-offs} between load balancing and scheduling overhead.

\noindent\textbf{SimGrid simulation toolkit.}
\simgrid{}~\cite{SimGrid} is a \mbox{well-known} toolkit based on \mbox{event-based} simulation. 
It supports the development of parallel and distributed applications in homogeneous/heterogeneous parallel and distributed environments. 
\simgrid{} has been selected for the current work due to its reliability and its active support in the community.
\simgrid{} has three main interfaces: \msg{}, \simdag{}, and \smpi{}.
\msg{} is used for the simulation of independent tasks.
\simdag{} supports simulation of tasks that have dependencies and represented as directed acyclic graphs (DAGs). 
\smpi{} is designed to simulate applications written using the message passing interface~\mbox{(MPI)}.

\noindent\textbf{Related work on DLS in simulation.}
Two interfaces of \simgrid{}, \msg{} and \simdag{}, were used to implement various DLS techniques. 
For instance, eight DLS techniques were implemented using the \msg{} interface \aliA{in the literature~\cite{mahad}}: five \mbox{non-adaptive}, SS, FSC~\cite{FSC}, GSS~\cite{GSS}, FAC~\cite{FAC}, and weighted factoring~(WF)\cite{WF}, and  three adaptive techniques, adaptive factoring~(AF)~\cite{AF}, adaptive weighted factoring~(\mbox{AWF-B} and \mbox{AWF-C})~\cite{AWFBC}.
The weak scalability of these DLS techniques \aliA{was} assessed in the presence of certain load imbalance sources (algorithmic and systemic).
The robustness of the same DLS techniques implemented using \msg{} was \aliA{also} studied~\cite{nitin}.
Moreover, the resilience of these DLS techniques on a heterogeneous computing system was studied using the \msg{} interface~\cite{dlsmsg}.
Another research effort used the \msg{} interface to reproduce certain experiments of DLS techniques~\cite{hoffeins2017examining}.  
\aliA{Therein, a successful reproduction of the past DLS experiments was presented. 
The results were compared to the experiments from the past to verify the implementation of the DLS techniques.} 
A similar approach of verifying the implementation of certain DLS techniques via reproduction was proposed using the \simdag{} interface~\cite{Ali_HPCC:17}. 
\aliA{The present work aims to assess the usefulness of these two \simgrid{} interfaces for achieving realistic simulations of scientific applications scheduled using the DLS implemented in \simgrid{}.}

\noindent\textbf{Related work on performance prediction.} A method was introduced for predicting the performance of dynamic load balancing techniques of geophysics applications using \simgrid{}~\cite{tesser2017}. 
In several \ahmed{g}eophysics applications, the level of the over-decomposition of the input domain and the chosen load balancing technique are key factors to achieve the desired performance.
However, the evaluation process of the possible combinations of these two elements is not feasible in many cases \ahmed{due to its} time and resource consuming process. 
Therefore, the \mbox{SMPI} interface of the \mbox{SimGrid}~\cite{SimGrid} simulation toolkit is extended to support the \mbox{Charm++'s} adaptive message passing interface~\mbox{(AMPI)}~\cite{AMPI}. 
This extension is referred to \mbox{SAMPI}.
The main idea is to port any MPI application to the AMPI library, then for once, the AMPI application is executed to obtain a time-independent trace~\cite{TiT}.
The \mbox{SAMPI} replayed the acquired trace using different load balancing techniques. 
The time needed for this replay is small compared to executing the real application.
The accuracy of this replay is subject to the accuracy of the hybrid flow-level network models of \mbox{SimGrid}~\cite{SimGrid:Network}.
This method relies on porting applications to the \mbox{AMPI} library which may require certain developing efforts. 
\mbox{AMPI} uses the concept of processor virtualization that is argued in~\cite{huang2006performance} to have no overhead.
However, there is a broad range of scientific applications that directly implement DLS techniques to balance the execution load. 
The methodology presented in the present work addresses the same concerns as \ahmed{Tesser et al.}~\cite{tesser2017} given that the evaluation of different load balancing techniques is time- and resource-consuming.


\section{Selected Parallel Application and Parallel Computing Systems}
\label{sec:app}
%
%

In this section, the application of interest and the computing systems under test \ali{are} introduced.
\subsection{The parallel application - PSIA}  
The application \ali{considered} in this work is an application from \ali{the} computer vision domain, namely, the parallel spin-image algorithm (PSIA)~\cite{psia}. 
 The SIA is a \mbox{computationally-intensive} application. The core computation of the SIA is the generation of the~2D spin-images. The PSIA exploits the fact that spin-images generations are independent of each other. 
\aliA{The size of a single spin-image is small (200~bytes) and fits in the lower level~(L1) cache. 
Therefore, the memory subsystem has an impact on the application performance.}

\begin{algorithm}[h]
	\SetKwInOut{Input}{Inputs}
	\SetKwInOut{Output}{Output}
	adCalculateSpinImages (W, B, S, OP, M, spinImages, start, end)\;
	\Input{W: \mbox{image width}, B: \mbox{bin size}, S: \mbox{support angle}, \mbox{OP: list of oriented points}, \mbox{M: number of oriented points}, \mbox{spinImages: list of spin-images to be filled}}
	\For{	{\color{blue} imageCounter = start $\rightarrow$ end }}
	{ 
		P = OP[imageCounter]\;
		tempSpinImage[W, W]\;
		init(tempSpinImage)\;
		\For{j = 0 $\rightarrow$ $M$}
		{
			X = OP[j]\;
			$np_i$ = getNormal(P)\;
			$np_j$ = getNormal(X)\;
			\If{	{\color{blue} acos($np_i \cdot np_j$) $\le S$}}
			{
				$k$  =  $\Bigg \lceil$ $\cfrac{W/2 - np_i \cdot (X-P) }{B}$ $\Bigg \rceil$ \;
				\vspace{0.2cm}
				$l$ =  $\Bigg \lceil$  $\cfrac{ \sqrt{||X-P||^2 - (np_i\cdot(X-P))^2} }{B}$  $\Bigg \rceil$\;
				
				\If{ {\color{blue} 0 $\le$ k $\textless$ W and 0 $\le$ l $\textless$ W}}
				{ tempSpinImage[k, l]++\;	}
			}
		}
		add(spinImages, tempSpinImage)\;
	}
	\caption{ \mbox{Spin-image} calculation algorithm~\cite{Eleliemy:2017b}}
	\label{algo:PSIA}
\end{algorithm}

A pseudocode of the PSIA~\cite{Eleliemy:2017b} is described in Algorithm~\ref{algo:PSIA}. According to Algorithm 1, lines 10 and 13, the amount of computations to generate spin-images is data-dependent and not identical over all the spin-images generated from the same object. 
This introduces an algorithmic source of load imbalance among the parallel processes generating the spin-images. 
The performance of the PSIA has been previously enhanced by using non-adaptive DLS techniques to balance the load between the parallel processes~\cite{Eleliemy:2017b}. 
Using DLS improved the performance of the PSIA by a factor of~1.2 and~2 for homogeneous and heterogeneous computing systems.
The number of spin-images generated by each process is governed by the \texttt{start} and \texttt{end} variables in Algorithm~\ref{algo:PSIA}, line~2. 
These variables represent the lower and upper bound of the indices of the images generated by a process, and the difference between them represents the chunk size calculated by the selected DLS technique.   


\subsection{The parallel computing systems}
\paragraph*{The miniHPC system}
The miniHPC is a high performance computing cluster of the Department of Mathematics and Computer Science at University of Basel, Switzerland. It consists of 26~compute nodes, a login node, and a storage node.  
The miniHPC cluster has a theoretical peak performance of $30~\mathit{TFLOP/s}$. 
For the experimental studies in this work, $22$~dual-socket nodes are used. Each node has two Intel Broadwell CPUs.
The four remaining compute nodes have \mbox{standalone} Intel Xeon Phi processors.
The software and hardware characteristics of the \emph{Broadwell partition} of the miniHPC system are listed in Table~\ref{tb:hpcs}.

\paragraph*{The Taurus system}
Taurus is a Bull HPC system at the Technische Universit{\"a}t Dresden, Germany. 
It comprises~$2,085$ nodes with a total theoretical peak performance of~$2,087~\mathit{TFLOP/s}$. 
For the experimental studies in this work, $22$~dual-socket Intel Broadwell nodes are used. 
The software and hardware characteristics of the \emph{Broadwell partition} of Taurus are listed in Table~\ref{tb:hpcs}.
\begin{table}[]
	\centering
	\caption{The characteristics of the HPC systems}
	\label{tb:hpcs}
		{\renewcommand{\arraystretch}{1.2}
	\begin{tabular}{l|l|l}
		\textbf{Parameter}         & \begin{tabular}[c]{@{}l@{}} \textbf{Broadwell partition} \\\textbf{of miniHPC}\end{tabular} & \begin{tabular}[c]{@{}l@{}} \textbf{Broadwell partition} \\\textbf{of Taurus}  \end{tabular}  \\ \hline
		Operating system  & \begin{tabular}[c]{@{}l@{}} CentOS Linux \\release $7.2.1511$  \end{tabular} &  \begin{tabular}[c]{@{}l@{}}  Red Hat Enterprise Linux \\ Server release 6.9 \end{tabular}       \\ \cline{2-3} 
		Job scheduler &     Slurm  v. $17.02.7$ &  Slurm v. $16.05.7$      \\ \cline{2-3} 
		MPI   & \begin{tabular}[c]{@{}l@{}}  Intel MPI \\  v. $2017$ update $1$\end{tabular}      &      \begin{tabular}[c]{@{}l@{}}  Intel MPI \\  v. $2017$ update $1$\end{tabular}    \\ \cline{2-3}  
 		File system       &          NFS4        &          NFS4       \\ \cline{2-3} 
		Number of nodes   &     $22$             &    32    \\ \cline{2-3} 
		Processor         &   \begin{tabular}[c]{@{}l@{}}  Intel Xeon\\  E5-2640 v4  \end{tabular}  &  \begin{tabular}[c]{@{}l@{}} Intel Xeon \\E5-2680 v4 \end{tabular}   \\ \cline{2-3} 
		Number of sockets &      $2$            &        $2$         \\ \cline{2-3} 
		Cores per socket  &    $10$              &       $14$          \\ \cline{2-3} 
		Hyper-threading   &       enabled           &     disabled            \\ \cline{2-3}  
		Operating frequency    &     $2.4~$-- $3.4~\mathit{GHz}$          &  $2.4~\mathit{GHz}$       \\ \cline{2-3} 
		 \begin{tabular}[c]{@{}l@{}}  Peak performance\\ per core  \end{tabular}  & $38.4$ -- $54.4~\mathit{GFLOP/s}$& $38.4~\mathit{GFLOP/s}$ \\ \cline{2-3} 
		L1 cache          &  $32~\mathit{KB}$ per core    &        $32~\mathit{KB}$  per core       \\ \cline{2-3} 
		L2 cache          &      $256~\mathit{KB}$ per core            &   $256~\mathit{KB}$ per core            \\ \cline{2-3} 
		L3 cache          &            $25~\mathit{MB}$ per socket      &    $35~\mathit{MB}$ per socket     \\ \cline{2-3} 
		RAM   &         $64~\mathit{GB}$ per node         &  $64~\mathit{GB}$ per node               \\ \cline{2-3} 
		Topology          &          non-blocking fat tree        &   non-blocking fat tree           \\ \cline{2-3} 
		Interconnection   &   Intel Omni-Path    &  Inifiniband FDR    \\ \cline{2-3} 
		Bandwidth         &        $100~\mathit{Gbit/s}$       &   $54.4~\mathit{Gbit/s}$      \\ \cline{2-3} 
		Latency           &          $100~\mathit{ns}$        &            $700~\mathit{ns}$    \\
	\end{tabular}
}
\end{table}

\subsection{Performance characterization of PSIA }
To test the application performance on both HPC platforms \ali{from Table~\ref{tb:hpcs}}, the application is executed to generate 400,000 spin-images from the Ramesses object~\cite{dataset} using 352 processes on 22 compute nodes.
The application is configured and bounded to use 16~cores on each node, to leave the rest of the cores for the operating system and other \mbox{system-related} processes. Each \ali{MPI} process is pinned to a single core \ali{among the} cores \ali{of} the two available processor sockets (16~processes on two sockets, 8~per socket) to \ali{uniformly} scatter the processes among the two non-uniform memory access (NUMA) domains and to avoid memory contention on a single \ali{NUMA} domain.
\ali{Each execution is repeated 20~times to obtain representative results.}
The \emph{coefficient of variation} (c.o.v.) of the processes finishing times of the parallel loop is used as a measure of the load imbalance~\cite{FAC}. The c.o.v. is defined as 
\begin{equation}
\label{eq:cov}
c.o.v. = \frac{\sigma}{\mu},
\end{equation}
where $\sigma$ is the standard deviation of the processes' parallel loop finishing times and $\mu$ is \ali{their} average. The c.o.v. is unitless and is bounded by 
\begin{equation}
\label{eq:bounds_cov}
0 \le c.o.v. \le \sqrt{(P - 1)},
\end{equation}
where $P$ is the number of processes~\cite{le2010performance}.
\ali{To quantify load imbalance in PSIA,} the finishing times of the processes are measured and the c.o.v. of these times is calculated. 

The results of \ali{executing PSIA with} the two load balancing extremes, STATIC and SS, are shown in Table~\ref{tb:cov_dls}. 
\aliA{The application execution time is denoted by $T_{par}$ and the parallel loop execution time is denoted by $T^{loop}_{par}$ . }
The results show that using SS achieved a balanced load execution with \ali{a} c.o.v. of~\aliA{0.003}  compared to~\aliA{0.022} with STATIC on miniHPC. 
Even though the c.o.v. values achieved by SS and STATIC are relatively small, the c.o.v. value achieved by SS is one order of magnitude \ali{lower} than that of STATIC, which indicates an improved load balance.
\aliA{Similarly}, the load balancing using SS on Taurus achieved a better performance than \aliA{with}STATIC.

\begin{table}[h]
	\centering
	\caption{The performance of the PSIA \ali{using STATIC and SS scheduling}. PSIA is configured to run with~352 processes to generate 400,000~spin-images. The median of 20~\ali{repetitions} of each experiment is reported.}
	\label{tb:cov_dls}
		{\renewcommand{\arraystretch}{1.2}
	\begin{tabular}{l|r|r|r|r}
		\textbf{HPC system}                                                              & \multicolumn{2}{c|}{\textbf{miniHPC}} & \multicolumn{2}{c}{\textbf{Taurus}} \\ \hline
		\begin{tabular}[c]{@{}l@{}}Loop scheduling\end{tabular}     & \multicolumn{1}{c|}{STATIC}         & \multicolumn{1}{c|}{SS}         & \multicolumn{1}{c|}{STATIC}         & \multicolumn{1}{c}{SS}         \\ \hline
			$T_{par}$ (s)                 &      \aliA{ $113.935$ }    &   \aliA{ $110.111$}    &   $181.516$   &  $177.890$    \\ 
		$T_{par}^{loop}$ (s)  &      \aliA{ $109.061$ }    &    \aliA{ $106.078$ }     &  $174.702$      &  $172.357$  \\
	
		C.o.v.  &$0.022$         &        $0.003$    &  $0.030$     &  $ 0.005$   \\ 
	\end{tabular}
}
\end{table}

\section{Methodology for \\Experimental Verification and Analysis }
\label{sec:method}
In this section, the proposed approach to analyze the application performance on different platforms is presented. In addition, this section describes how to extract \ali{the} application and the platform characteristics and \ali{how to} represent them in simulation. The experimental verification methodology of the simulative execution of the application is described next.
\subsection{Extraction of application and platform characteristics}
\label{subsec:extract}
\ali{Characterizing} the \ali{behavior} of a parallel application on an HPC system can be challenging as it involves the representation of two major components that contribute to \ali{its} performance: (1)~The application representation; and (2)~The HPC system. Using the approach introduced \ali{in an earlier work}~\cite{Ali_SC:17}, the representation of the computing system can be verified \ali{via} a separation of the application representation by using the \smpi{} interface. 
The \smpi{} interface simulates the execution of native message passing interface (MPI) codes on a simulated computing platform. 
Both the native and simulative executions using \smpi{} share the application's native code. The difference between the native execution and the simulative \mbox{\smpi{}-based} execution is the computing system component. 
\ali{The} representation of the computing system can be verified by comparing the native and \smpi{} simulative performance results.
The \smpi{} simulation produces a special type of text-based execution trace called time independent trace (TiT)~\cite{TiT}. 
The TiT contains a trace of the application execution as a series of computation and communication events, with their amounts specified in FLOP and bytes, respectively. 
\ali{The} TiT is used to understand the application flow and to represent the application in \ali{the} \msg{}/\simdag{} interfaces. 
The same computing system representation used earlier in the \smpi{} simulation is used for \ali{the} \msg{}/\simdag{} simulations. 
The performance results of the \msg{}/\simdag{} simulations are compared to the native execution results to verify the application representation. 

The amount of work contained in each iteration of the loop is measured using PAPI~\cite{papi}. 
The FLOP count \ali{obtained with} PAPI is used to represent the amount of work in each iteration in \msg{}/\simdag{}. 
\aliA{The core speed needs to be estimated to obtain more accurate simulation results, due to the fact that the application does not execute at the theoretical peak performance.}
The core speed is calculated by measuring the loop execution time in a sequential run to avoid any parallelization or communication overhead. The sum of the total number of FLOP in all iterations is divided by the measured loop execution time to estimate the core processing speed. 
This core speed is used in the \simgrid{} \texttt{platform file} to represent the computing system core speed in processing the application loop iterations. 
%
%

%

\subsection{Employing DLS in native and simulative executions }
In a recent work~\cite{Eleliemy:2017b}, DLS has been used to balance the load of PSIA application executing on homogeneous and heterogeneous computing platforms. 
\ali{The performance of the PSIA was studied using strong and weak scalability on more than $300$~heterogeneous cores. 
Using DLS enhanced the performance of PSIA by a factor of $1.2$~and $2$~compared to STATIC, for homogeneous and heterogeneous platforms, respectively.}

\subsection{Simulation of PSIA}
\label{sec:simulation_apps}
\noindent\textbf{MSG simulation.}
\fk{
The \msg{} module implements a master-worker execution model. 
Two-sided communication is used for information exchange between master and workers. 
This characteristic fits perfectly the demands of studying scheduling algorithms using a central entity for coordination of the work distribution. 
However, in this work an approach that coordinates the distribution of work via common state information held in memory is investigated. 
Instead of using a central entity and two-sided communication, the access to the state information needed for scheduling decisions is achieved via one-sided communication. 
Therefore, the master in the \msg{} representation of the PSIA application represents this state information. 
Whenever a worker is available or becomes idle, it makes a request to the master before computing the next chunk size. 
This request represents the remote memory access. 
}\\

\noindent\textbf{SimDag simulation.}
In \simdag{}, the applications are represented as directed acyclic graph~(DAG) of tasks. Dependencies can be added between tasks to represent execution precedence. 
Tasks can be computation tasks or communication tasks. To represent the PSIA application, each loop iteration is represented as a computation task. 
The amount of work in a \ali{computational} task is equal to the FLOP counted by PAPI for the corresponding loop iteration. 
The FLOP count of all iterations is read from a file to create computational tasks \ali{in simulation} that represent loop iterations. 
Whenever a process is \ali{available}, the scheduler calculates a chunk size and allocates it to this process. 
A computation and a communication tasks are created at each scheduling step to represent the scheduling overhead in calculating a chunk size and the \ali{communication with} the requesting process. 
The amount of work contained in the \ali{tasks denoting the} computation scheduling overhead is acquired \ali{with} PAPI to count the FLOP in the functions that calculate and assign chunks of work in \ali{the} native code. 
The amount of communication in the \ali{tasks denoting the} scheduling overhead is equal to four bytes, which represents \ali{the} communication of one integer (chunk size). 
The DLS implementation in \simgrid{} \ali{was} verified in previous work~\cite{Ali_HPCC:17}.\\

\noindent\textbf{SimGrid platform files.}
The characteristics of the computing platform \ali{are} provided to the simulator using an \texttt{XML} file, called the \texttt{platform file}. 
Each computing node is represented a host in the \texttt{platform file}. 
The number of cores \ali{of} a host is equal to the number of cores in \ali{a node of} the computing platform (20 \aliA{cores per node} in miniHPC and 28 \aliA{cores per node} in Taurus). 
The core speed \ali{calculation is} described in Section~\ref{subsec:extract} for both platforms. 
The core speeds are found to be $0.705$~GFLOP/s and $0.439$~GFLOP/s for miniHPC and Taurus, respectively. 
Both platforms use the \ali{non-blocking} fat tree topology with different parameters to describe the number of \ali{fat tree} levels, nodes, and links. 
The link bandwidth and latency are also specified in the \texttt{platform files} of the two systems.
\aliA{The \simgrid-based calibration procedure~\cite{simgrid_calib} is used to calibrate the representations of both platforms to better adjust the network bandwidth and latency in their respective \texttt{platform files}.}

\subsection{Experimental verification and analysis}
\begin{figure}[]
	\centering
	\includegraphics[clip, trim=0.5cm 0cm 1cm 0.5cm,scale=0.6]{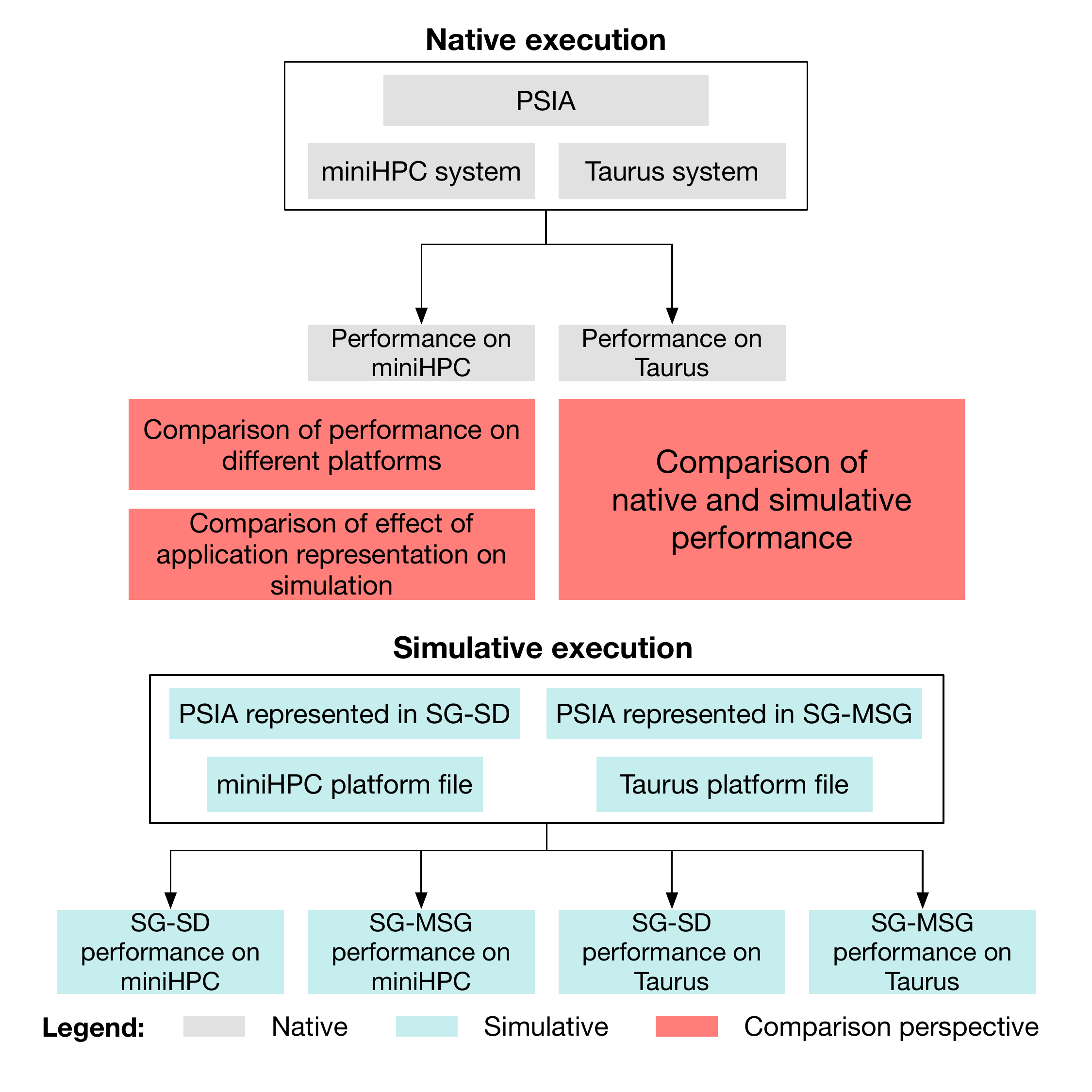}
	\caption{Experimental verification and analysis methodology.}
	\label{fig:comapp}
\end{figure}

Three perspectives of comparisons are \ali{taken} in this work to analyze the performance of the PSIA in native and simulative executions as depicted in \figurename{~\ref{fig:comapp}}. 
Through the first perspective, the native performance of the PSIA on both HPC platforms \ali{is} compared. 
This comparison allows the analysis of how the native platform characteristics \ali{influence} the performance of the application \ali{for the DLS techniques considered}. 
In the second perspective, the performance of the simulative executions from \msg{} and \simdag{} \ali{is} compared to evaluate how the application representation (\ali{using data or task parallelism}) can affect the simulative results. 
Through the third perspective, the native and simulative results are compared to \ali{answer: \emph{`How realistic are \aliA{the} simulations of executions of scientific applications using DLS on HPC platforms?'}.}
The first and the second comparisons are prerequisites for the third comparison, to understand the application characteristics (first) and to evaluate the effect of the application representation on the simulative results (second).

To compare which interface \ali{better predicts} the performance of the application of interest, the performance of the native \ali{and the simulated application} from the \msg{} simulation and the \simdag{} simulation \ali{is compared}. 
The percent error $\%E$ between native \ali{performance}~($T_{nat}$) and simulative performance~($T_{sim}$) is calculated as 
 \begin{equation}
 \label{eq:reldiff}
 \%E = \left(1- \frac{T_{sim}}{T_{nat}}\right) \times 100.
 \end{equation}
The \ali{percent errors} between native and simulative performance are compared to answer the following \ali{additional} questions: \textit{(1)~Which interface simulates the application performance with a \ali{reduced} $\%E$ for the same computing system? (2)~Which interface simulates the application performance with a minimum change in the $\%E$ in predicting the performance on the two machines (miniHPC and Taurus)?} The answers to these questions are essential to understand the accuracy of the simulation performance predictions on different systems. \ali{These} results will guide future decisions regarding which interface to use for a given application and computing system and on \ali{the expected accuracy of the simulation results}.


%
%
\section{Results of Executing PSIA using DLS}
\label{sec:predict}
The results of native executions of the PSIA on the two considered HPC systems, miniHPC and Taurus, are presented in this section. The simulative executions results of the PSIA using the two simulation interfaces, \msg{} and \simdag{}, are also illustrated and compared to the native executions results. A discussion on the results of the comparisons between the performance on the two HPC systems in native executions, the performance obtained from simulative executions using \msg{} and \simdag{}, and the percent error between the native and simulative executions is also included.

\subsection{Native execution results}
\begin{landscape}
\begin{figure*}[]
	\centering
	\subfloat[Application execution time (miniHPC)]{%
	\includegraphics[clip, trim=0cm 0cm 0cm 0cm,scale=0.3]{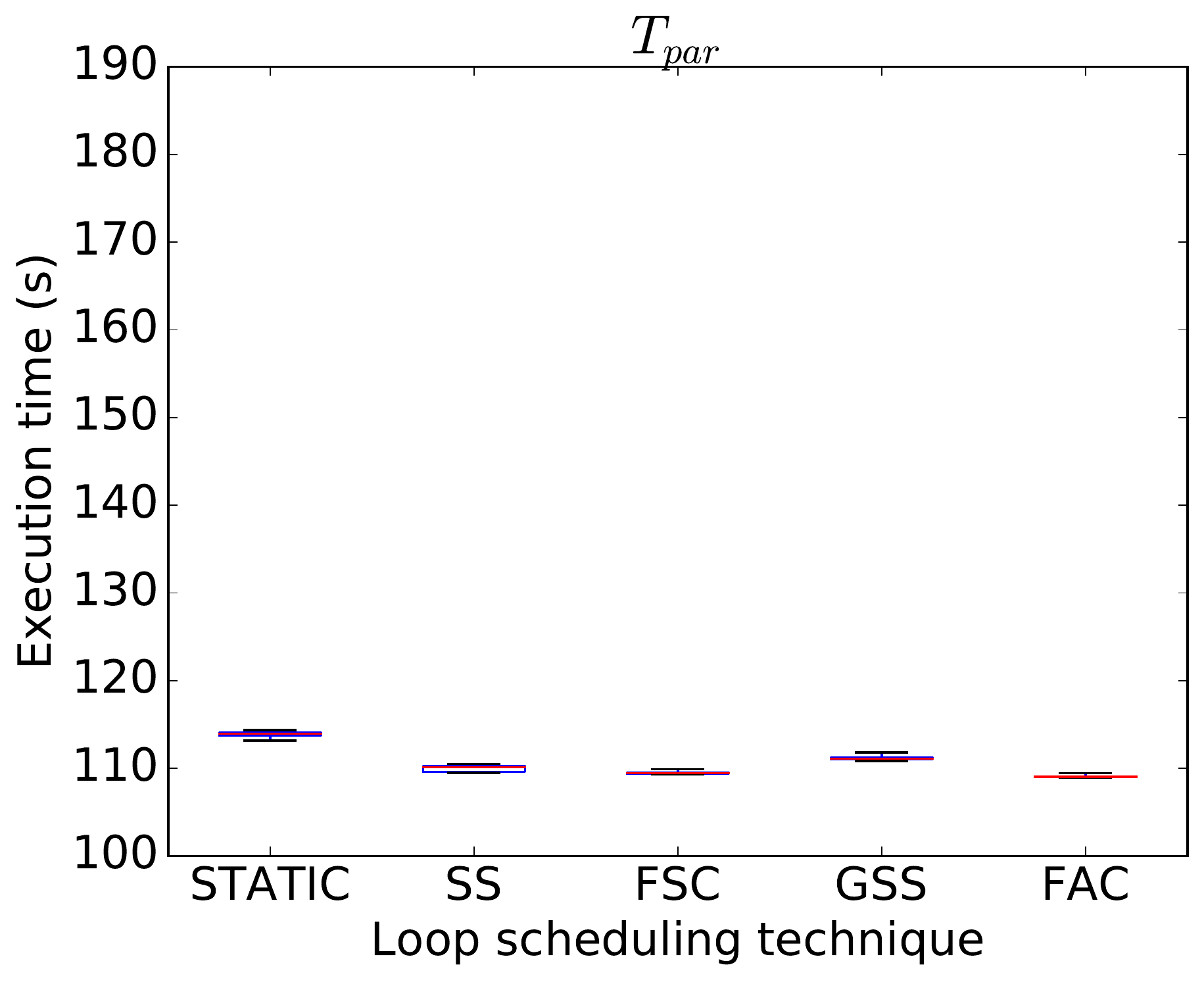}%
		\label{fig:native_entire_miniHPC}%
	} 
	\subfloat[Parallel loop execution time (miniHPC)]{%
		\includegraphics[clip, trim=0cm 0cm 0cm 0cm,scale=0.3]{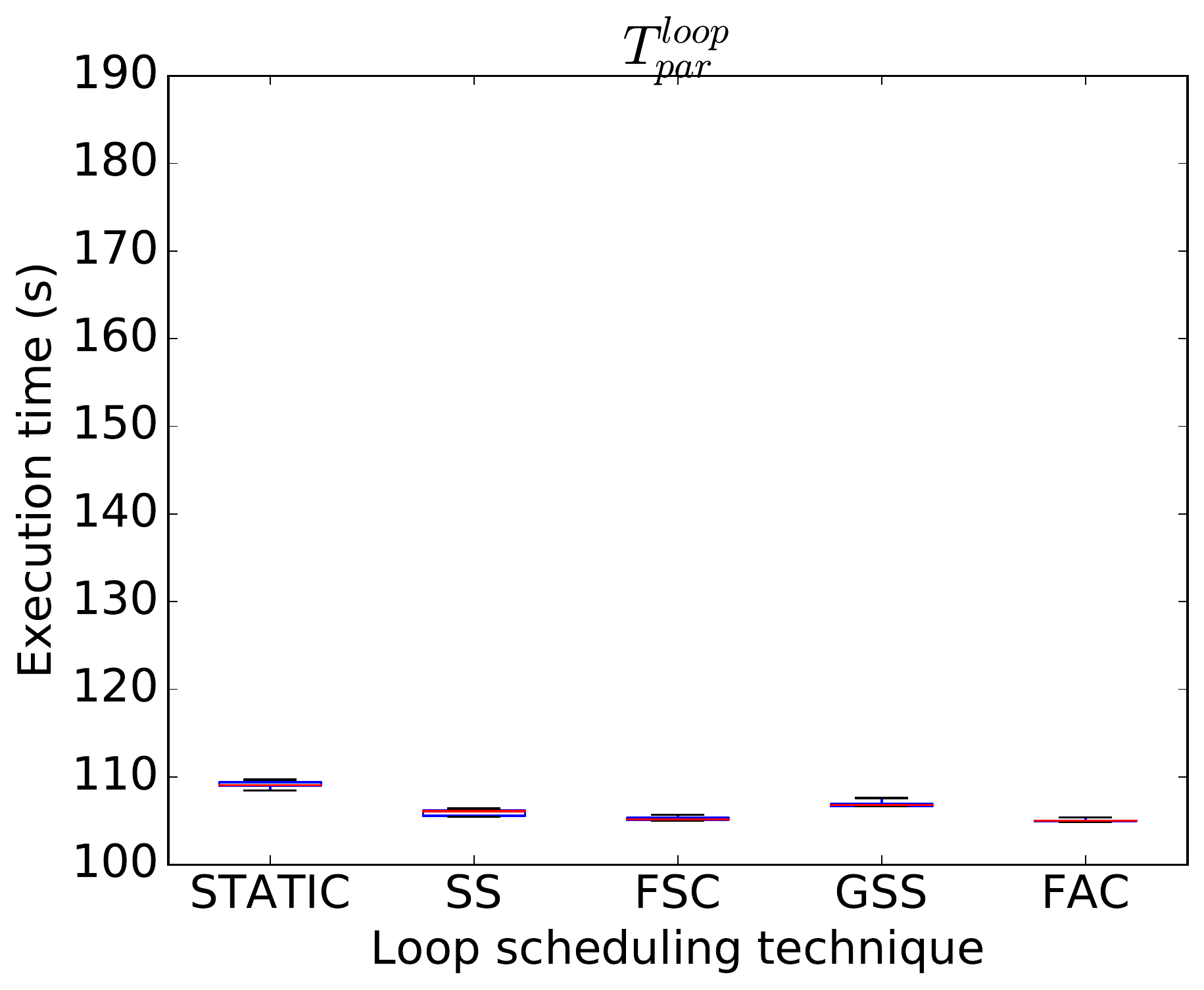}%
		\label{fig:native_tpar_miniHPC}%
	} 
	\subfloat[C.o.v. of processes parallel loop finishing times (miniHPC)]{%
		\includegraphics[clip, trim=0cm 0cm 0cm 0cm,scale=0.3]{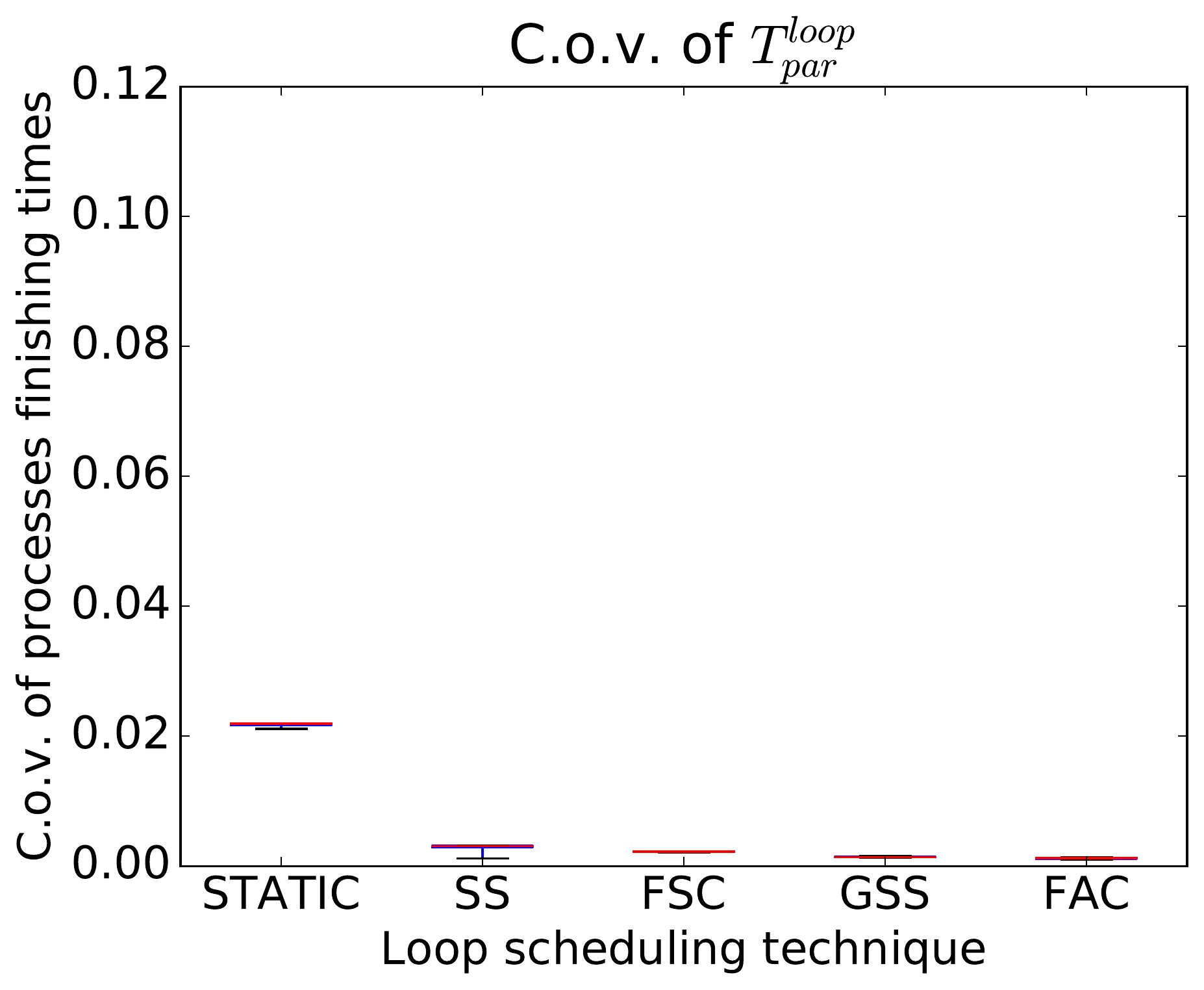}%
		\label{fig:native_cov_miniHPC}%
	} 
	\\ 
	\subfloat[Appplication execution time (Taurus)]{%
	\includegraphics[clip, trim=0cm 0cm 0cm 0cm,scale=0.3]{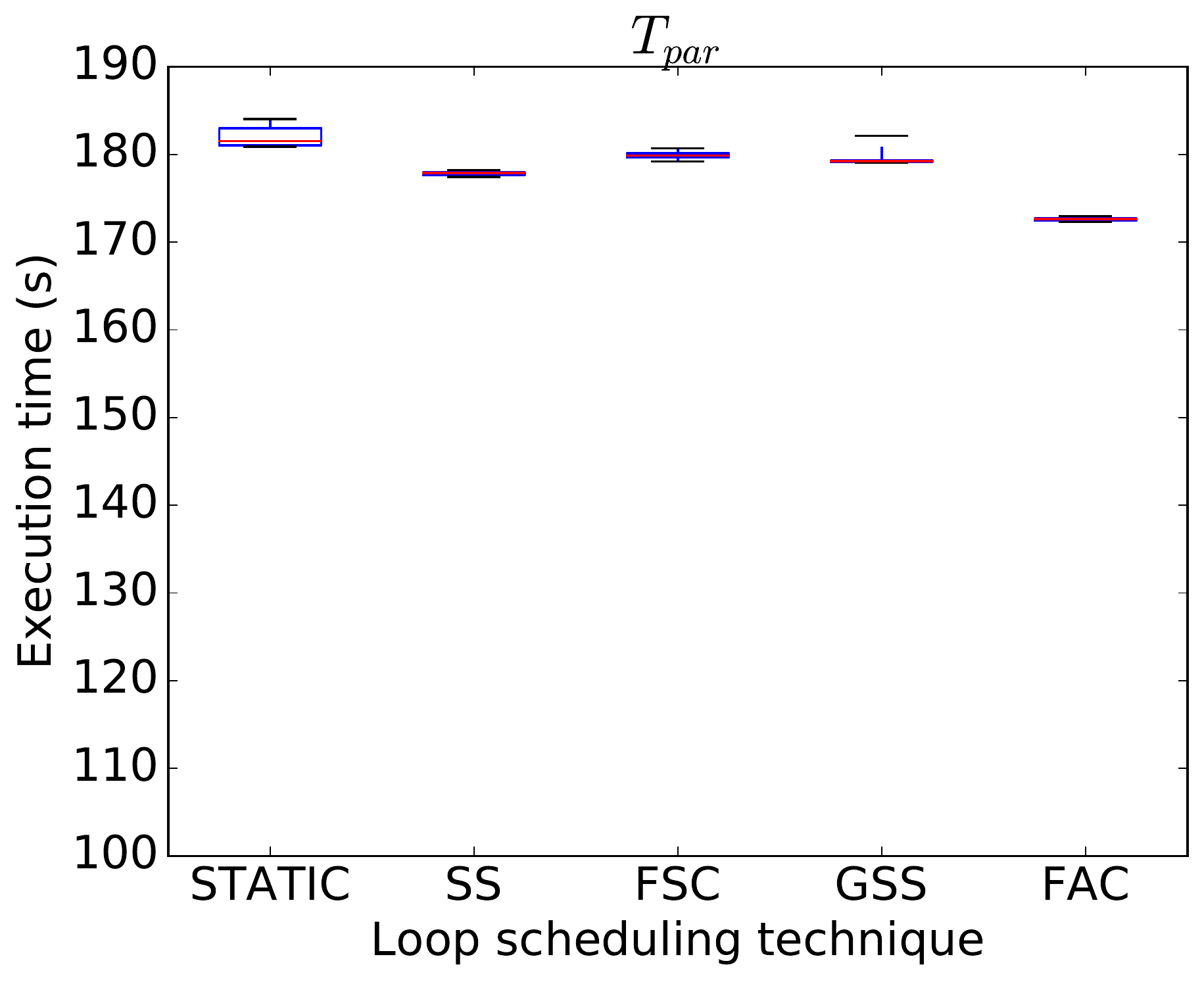}%
		\label{fig:native_entire_Taurus}%
	}
	\subfloat[Parallel loop execution time (Taurus)]{%
			\includegraphics[clip, trim=0cm 0cm 0cm 0cm,scale=0.3]{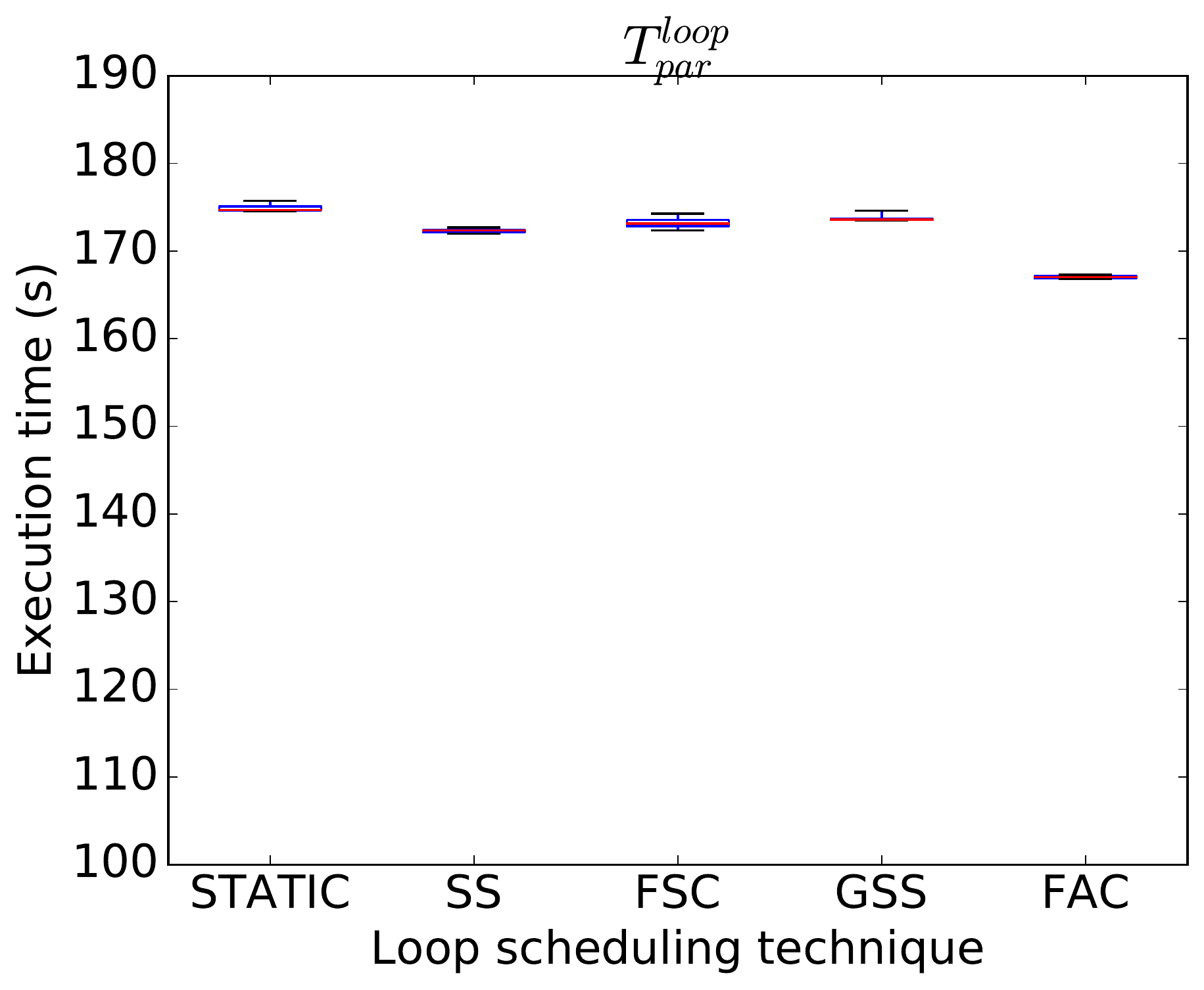}%
		\label{fig:native_tpar_Taurus}%
	}
	\subfloat[C.o.v. of processes parallel loop finishing times (Taurus)]{%
	\includegraphics[clip, trim=0cm 0cm 0cm 0cm,scale=0.3]{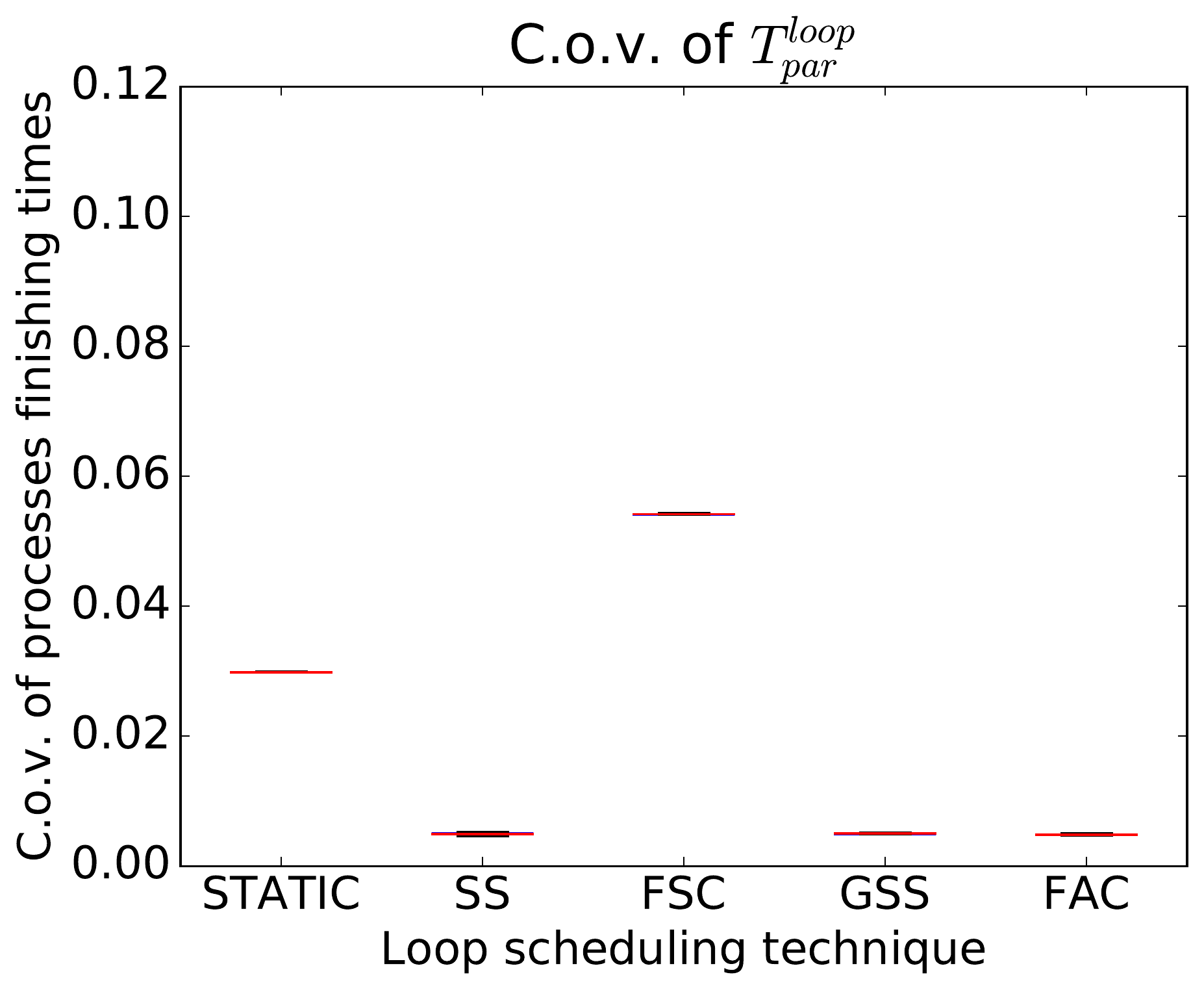}%
		\label{fig:native_cov_Taurus}%
	}
	\caption{{\footnotesize\emph{Native} performance of the PSIA using DLS on miniHPC and Taurus. PSIA is configured to generate 400,000~\mbox{spin-images} from the Ramesses object~\cite{dataset}. The PSIA is executed using 352~processes, on 22~compute nodes, and 16~processes per node. The processes are distributed and binded to processor cores from the two processor sockets, 8~processes per socket. Experiments are repeated 20~times. The whiskers represent the maximum and the minimum values. The box represents the first and the third quartiles of the results data. The red line represents the median value.}}
	\label{fig:native}
\end{figure*}    
\end{landscape}

The results of the native execution of the PSIA on miniHPC and Taurus are depicted in \figurename{~\ref{fig:native}}.
The results in \figurename{~\ref{fig:native_cov_miniHPC}} and \figurename{~\ref{fig:native_cov_Taurus}} show that using SS, GSS, and FAC achieved a balanced load execution on both systems. 
The FSC failed to achieve a balanced execution \aliA{on Taurus} compared to the other DLS techniques. 
This may be due to a suboptimal estimation of the scheduling overhead, $h$, and the standard deviation of the loop iterations execution times, $\sigma$, that are needed by the FSC to properly calculate the chunk sizes. 
Even though SS, GSS, and FAC achieved a balanced load execution and very small values of c.o.v., FAC outperformed all other scheduling techniques considered in this work. 
This is due to the better load balancing of FAC and its lower scheduling overhead compared to SS and GSS. 

\subsection{Simulative execution results}
\begin{landscape}
\begin{figure*}[]
	\centering
	\subfloat[Application execution time (miniHPC)]{%
		\includegraphics[clip, trim=0cm 0cm 0cm 0cm,scale=0.3]{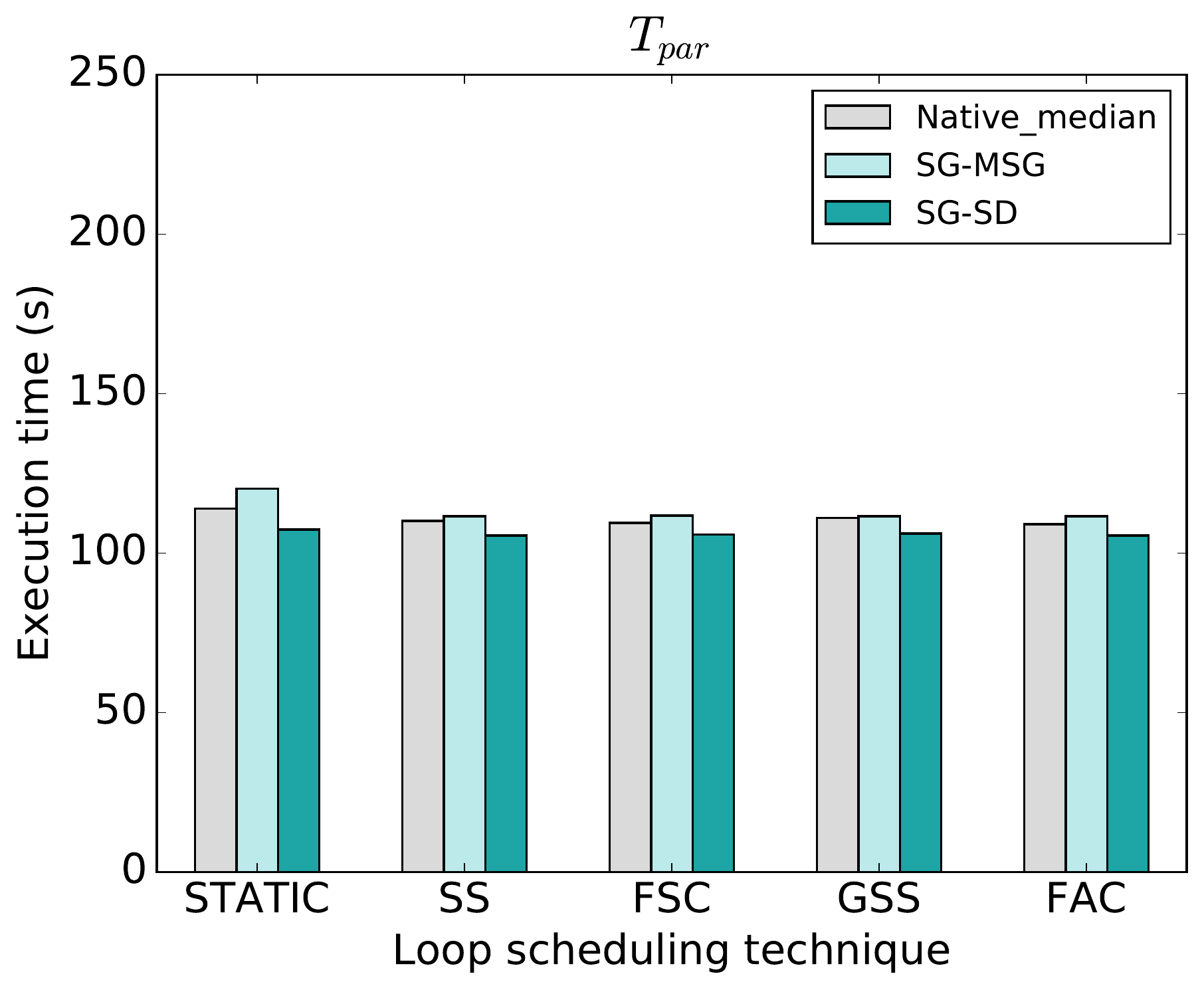}%
		\label{fig:sim_entire_miniHPC}%
	} 
	\subfloat[Parallel loop execution time (miniHPC)]{%
		\includegraphics[clip, trim=0cm 0cm 0cm 0cm,scale=0.3]{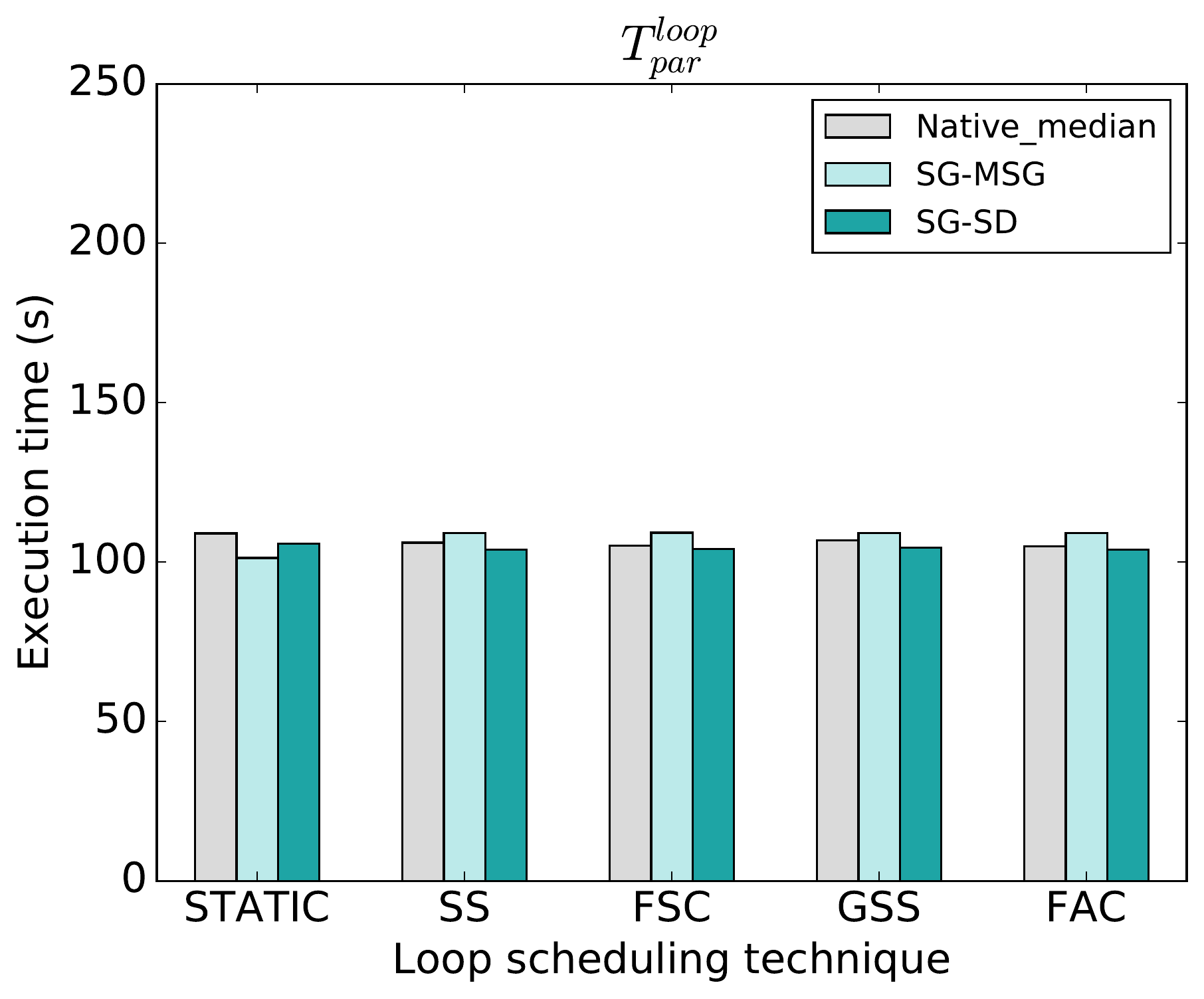}%
		\label{fig:sim_tpar_miniHPC}%
	} 
	\subfloat[C.o.v. of processes parallel loop finishing times (miniHPC)]{%
		\includegraphics[clip, trim=0cm 0cm 0cm 0cm,scale=0.3]{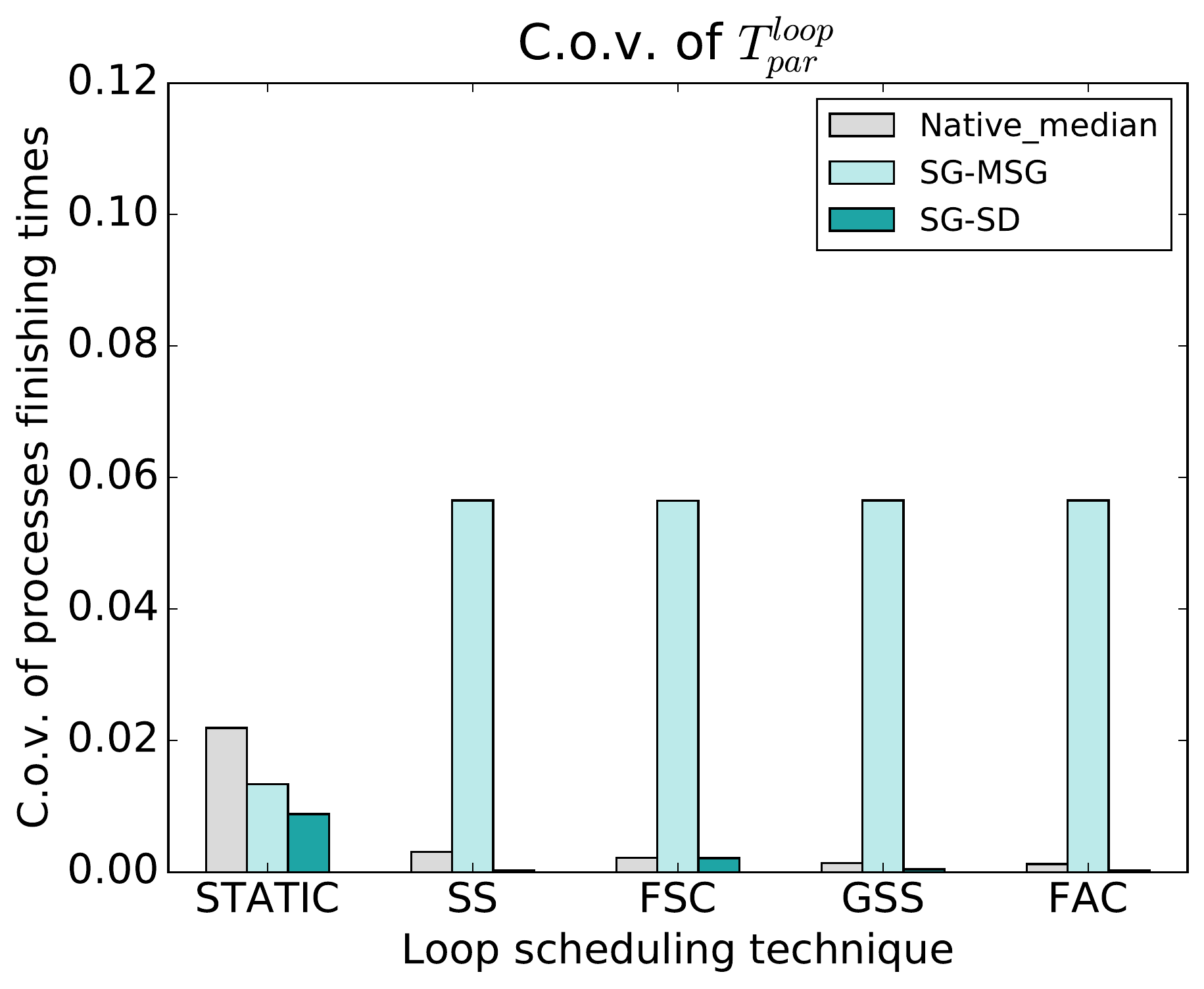}%
		\label{fig:sim_cov_miniHPC}%
	} 
	\\ 
	\subfloat[Application execution time (Taurus)]{%
		\includegraphics[clip, trim=0cm 0cm 0cm 0cm,scale=0.3]{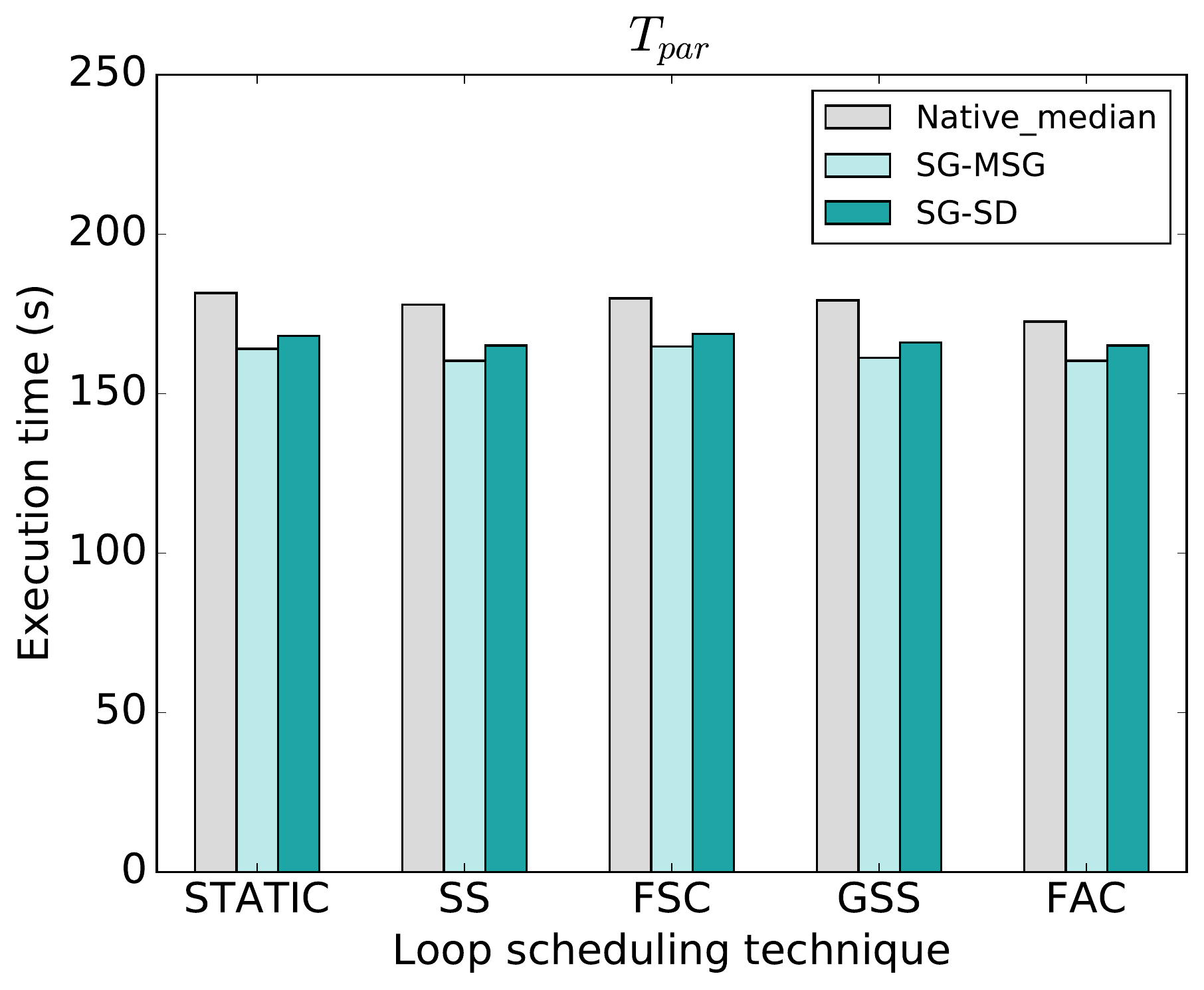}%
		\label{fig:sim_entire_Taurus}%
	}
	\subfloat[Parallel loop execution time (Taurus)]{%
		\includegraphics[clip, trim=0cm 0cm 0cm 0cm,scale=0.3]{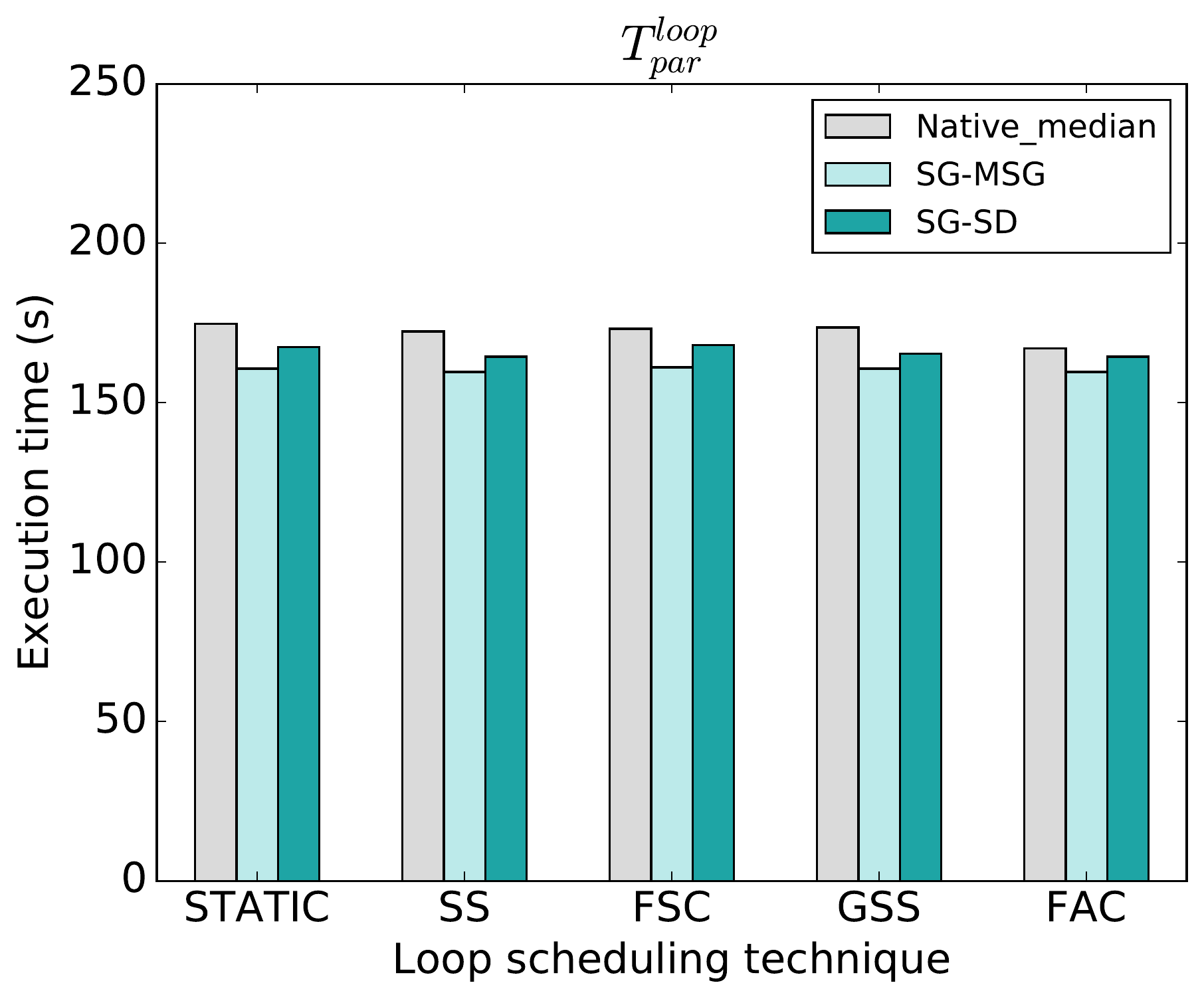}%
		\label{fig:sim_tpar_Taurus}%
	}
	\subfloat[C.o.v. of processes parallel loop finishing times (Taurus)]{%
		\includegraphics[clip, trim=0cm 0cm 0cm 0cm,scale=0.3]{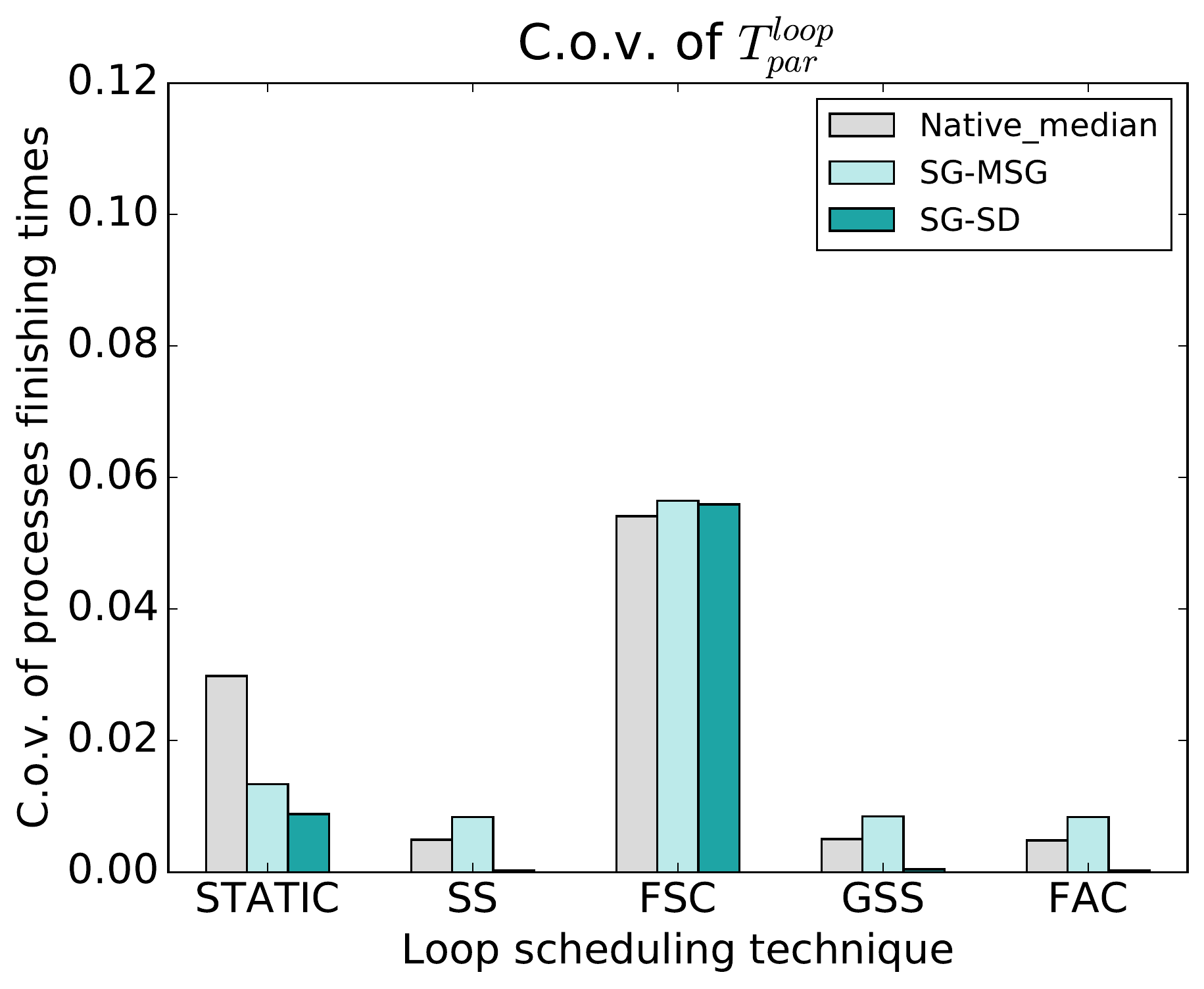}%
		\label{fig:sim_cov_Taurus}%
	}
	\caption{\emph{Simulative} performance of the PSIA  using DLS obtained using \msg{} and \simdag{}. Simulative execution results are compared \aliA{with} the median of \ali{20 repetitions of} the \emph{native} executions results.}
	\label{fig:sim}
\end{figure*}    
\end{landscape}

The results of the simulative executions with \msg{} and \simdag{} compared to the median of the native execution on both, miniHPC and Taurus, are shown in \figurename{~\ref{fig:sim}}. 
The results show that, in general, the simulative execution tends to underestimate the execution times. 
Both simulations using \msg{} and \simdag{} captured the large value of the c.o.v. in the case of FSC on Taurus, as can be seen in \figurename{~\ref{fig:sim_cov_Taurus}}. 
The \msg{} simulation tends to overestimate the c.o.v. values, especially on miniHPC, whereas \simdag{} simulation underestimates the c.o.v. values in most cases. 
The results of both simulation interfaces (\msg{} and \simdag{}) tend to underestimate the execution time \aliA{on Taurus} for scheduling techniques that incur high overhead, such as, SS, as the scheduling overhead is not fully captured by both simulations. 
For example, \msg{} only accounts for the messages to send the chunk size whereas \simdag{} considers the FLOP count of the chunk calculation and the messages to send the chunk size.
\aliA{The \msg{} simulation tends to overestimate the execution time on miniHPC in most cases, contrary to the \simdag{} simulation, which always underestimates the execution time on miniHPC and on Taurus.}

For the execution on miniHPC, \aliA{both the \msg{} and the \simdag{} simulations correctly predict that STATIC results in the worst performance and that FAC outperforms all other techniques, similar to the native execution.}
For the execution on Taurus, both simulators correctly predict that FAC outperforms other loop scheduling techniques. 
Both simulations incorrectly predict that the worst performance occurs with FSC, instead of STATIC as the native execution.
Comparing the two simulative execution results obtained with \msg{} \ali{against} results obtained with \simdag{} from the perspective of the percent errors calculated between the native and simulative parallel loop execution times in Table~\ref{tbl:rel_diff}, one can see that \aliA{\msg{} simulation tends to overestimate the execution time on miniHPC and underestimate the execution time on Taurus.} 
\aliA{Inspecting the percent errors in the case of the \simdag{} execution, one can see that it always underestimates the execution time on both HPC systems.}
The median of the $\%E$ of the \msg{} simulations is \aliA{$−2.89\%$} compared to \aliA{$2.05\%$} with the \simdag{} simulation for the miniHPC execution. For the execution on Taurus, the medians of the $\%E$ are $7.37\%$ and $4.14\%$ for \msg{} and \simdag, respectively. These median $\%E$ values are considered small, and the larger values of the $\%E$ can be decreased in the future \ali{through} a better representation of the scheduling overhead. 

\begin{table}[]
	\centering
	\caption{The percent error $\%E$ between native and simulative parallel loop execution times of the PSIA on miniHPC and Taurus. The $\%E$ is calculated as Section~\ref{eq:reldiff}.}
	\label{tbl:rel_diff}
	{\renewcommand{\arraystretch}{1.2}
	\begin{tabular}{l|r|r|r|r}
		\multirow{2}{*}{\textbf{\begin{tabular}[c]{@{}l@{}}Loop scheduling\\ technique\end{tabular}}} & \multicolumn{2}{l|}{\textbf{\begin{tabular}[c]{@{}l@{}}\msg{} simulation of\end{tabular}}} & \multicolumn{2}{l}{\textbf{\begin{tabular}[c]{@{}l@{}}\simdag{} simulation of\end{tabular}}} \\ \cline{2-5} 
		& \multicolumn{1}{c|}{\textbf{miniHPC}}             & \multicolumn{1}{c|}{\textbf{Taurus}}             & \multicolumn{1}{c|}{\textbf{miniHPC}}               & \multicolumn{1}{c}{\textbf{Taurus}}              \\ \hline
		STATIC   &   \aliA{$7.14\%$}    &  $8.03\%$         & \aliA{$2.99\%$}    &  $4.14\%$  \\
		SS          &   \aliA{$-2.89\%$}    & $7.37\%$        & \aliA{$2.05\%$}     &  $4.58\%$   \\
		FSC        &   \aliA{$-3.83\%$}     & $6.96\%$       & \aliA{$0.95\%$}    &  $2.91\%$    \\
		GSS       &    \aliA{$-2.21\%$}      & $7.46\%$      & \aliA{$2.16\%$}     &   $4.72\%$   \\
		FAC       &    \aliA{$-3.93\%$}      & $4.42\%$      &  \aliA{$1.04\%$}   &   $1.55\%$                   \\    
	\end{tabular}
}
\end{table}

\subsection{Discussion}
The native and simulative experiments performed in this work and the analysis of their results have revealed certain key aspects. 

First, the application representation and the platform representation in simulation \emph{can not be decoupled}. 
One needs to take into consideration how the application  characteristics specified in the simulation interact with the represented computing platform. 
\aliA{For example, the processor core speed is measured as described in Section~\ref{subsec:extract} to achieve more accurate simulation results.}

\aliA{Second, representing the computational effort in an application using FLOP count was used as it represents the amount of work regardless of the platform computing speed as opposed to time measurements. 
Also, FLOP count can be accurately measured even at the fine-grained loop iterations.}
Measuring the execution times of loop iterations to represent the computational effort in the loop iterations was also not very successful. 
The \mbox{time-based} measurements could not capture the dynamic behavior of the application and is affected by the measurement process. 
The results of simulations performed using time-based measurements to represent the application can be found online~\cite{repro}.
\aliA{The time measurement is used at the gross grain of the loop execution time to estimate the core speed as described in Section~\ref{subsec:extract}.}

Third, the comparison between native and simulative execution results confirms that a close agreement thereof is limited by the absence of modern CPU and memory models. 
\aliA{SimGrid uses a simple CPU model, where the computation time is equal to the computational effort in a loop iteration divided by the core speed~\cite{SimGrid}.}
The application and platform characteristics need to be aligned with the simulator and the subsystems models it offers. 
For example, to accurately predict the performance of a \mbox{memory-bound} application, a simulator that offers a precise memory model, in addition to other subsystems models, is required. 
Other simulators may provide more complex models and more accurate results, however, they may not be adequate for the purpose of studying scheduling techniques. 
Finally, the choice of the application model representation in simulation (using data or task parallelism) may affect the simulation results.

\section{Conclusion and Future Work}
\label{sec:con}
\fk{
In this work, a methodology is devised to answer the question \aliA{of} how realistic are \aliA{the} simulations of executions of scientific applications using DLS on HPC platforms, and involves \aliA{the experimental verification and the performance analysis of DLS in scientific applications.} 
\aliA{The answer to this question helps to eliminate the uncertainty regarding the performance results obtained via simulation.}
An approach is proposed to analyze the performance of an application on different platforms. 
\aliA{This work} described how to extract the application and the platform characteristics and \aliA{how} to represent them in simulation. 
Furthermore, the experimental verification methodology of the simulative execution of the application is explained. 
The proposed methodology is employed for a computer vision application executing using four DLS techniques on two different HPC platforms, both via native and simulative (using two SimGrid interfaces) experiments.
The evaluation and the analysis of the native and simulative results indicate that the accuracy of the simulative experiments is strongly influenced by the values obtained by the chosen approach used to extract the computational effort of the application (FLOP- or time-based), the choice of application model representation into simulation (data or task parallel), and the choice of HPC subsystem models available in the simulator (multi-core CPUs, memory hierarchy, and network topology). 
\aliA{The minimum percent error achieved between native and simulative experiments was $0.95\%$, while the maximum was $8.03\%$}
Further work remains for arriving at \aliA{an even} closer agreement between the native and simulative results via more precise representation of the application and system characteristics. 
\aliA{The study of the effect of memory system on the performance of scientific application would lead to a more accurate application representation, thus, closer simulative results to the native results.}
Furthermore, the presented methodology can be employed for other computationally-intensive, scientific applications \aliA{using adaptive DLS techniques,} with the goal to improve their performance on real HPC systems by selecting via simulation the best suited DLS. 
}


\section*{Acknowledgment}
\fk{
This work was in part supported by the Swiss National Science Foundation in the context of the ``Multi-level Scheduling in Large Scale High Performance Computers'' (MLS) grant number 169123 and the USA National Science Foundation under grant number NSF CGI-1034897.
}
 
\section{Reproduction of This Work}
\label{sec:reproduction}
Reproducibility of the execution of scientific applications on parallel and distributed computing systems is of a growing interest, underlying the trustworthiness of the experiments and the conclusions derived from experiments. 
In the following subsections, it is described how the source codes were compiled and how experiments were executed.

\subsection{Native experiments on miniHPC and Taurus}
The native PSIA code 
was compiled with the Intel MPI compiler version 2017 update 1 with -O0 compiler optimization level flag. 
For each of the loop scheduling techniques STATIC, SS, FSC, GSS, and FAC, $20$ runs were performed. 
Each run of the application was executed in a single job. 
Slurm was used for job scheduling. 
The application was executed using $352$ tasks (processes) on $22$ compute nodes, with $16$ processes per node. 
The Slurm \texttt{exclusive} flag was used to prevent the scheduler from assigning the same nodes to other jobs simultaneously and avoid interference.  

The Intel MPI \texttt{I\_MPI\_ASYNC\_PROGRESS} flag was set to speed up the execution of the one-sided communications. 
The ``Tag Matching Interface (TMI)'' MPI fabrics library was used on the miniHPC as it provides improved performance on the Intel Omni-Path interconnection fabric, whereas the ``Direct Access Programming Library (DAPL)'' fabrics library was used on the Taurus.
The \texttt{srun}-command was used to launch the application processes on the 22 nodes. 
Processes were pinned to cores from the two processor sockets using scatter strategy to balance the load among the two sockets.

\subsection{Simulative experiments using \msg{} and \simdag{}}
The SimGrid simulation framework version $3.16$ was compiled using Intel compiler 2017 update 1. 
The \msg{} and \simdag{} codes 
were compiled using the Intel C compiler (\texttt{icc}) with \texttt{-g -Wall} compilation flags. 

 \bibliographystyle{IEEEtran}
\bibliography{predicting}

\begin{thebibliography}{10}
\providecommand{\url}[1]{#1}
\csname url@samestyle\endcsname
\providecommand{\newblock}{\relax}
\providecommand{\bibinfo}[2]{#2}
\providecommand{\BIBentrySTDinterwordspacing}{\spaceskip=0pt\relax}
\providecommand{\BIBentryALTinterwordstretchfactor}{4}
\providecommand{\BIBentryALTinterwordspacing}{\spaceskip=\fontdimen2\font plus
\BIBentryALTinterwordstretchfactor\fontdimen3\font minus
  \fontdimen4\font\relax}
\providecommand{\BIBforeignlanguage}[2]{{%
\expandafter\ifx\csname l@#1\endcsname\relax
\typeout{** WARNING: IEEEtran.bst: No hyphenation pattern has been}%
\typeout{** loaded for the language `#1'. Using the pattern for}%
\typeout{** the default language instead.}%
\else
\language=\csname l@#1\endcsname
\fi
#2}}
\providecommand{\BIBdecl}{\relax}
\BIBdecl

\bibitem{Banicescu2005}
I.~Banicescu and R.~L. Cari{\~n}o, ``Addressing the stochastic nature of
  scientific computations via dynamic loop scheduling,'' \emph{{Electronic
  Transactions on Numerical Analysis (ETNA)}}, vol.~21, pp. 66--80, 2005.

\bibitem{psia}
A.~Eleliemy, M.~Fayze, R.~Mehmood, I.~Katib, and N.~Aljohani, ``{Loadbalancing
  on Parallel Heterogeneous Architectures: \mbox{Spin-image} Algorithm on CPU
  and MIC},'' in \emph{Proceedings of the 9th EUROSIM Congress on Modelling and
  Simulation}, September 2016, pp. 623--628.

\bibitem{johnson1999using}
A.~E. Johnson and M.~Hebert, ``{Using spin images for efficient object
  recognition in cluttered 3D scenes},'' \emph{{IEEE Transactions on Pattern
  Analysis and Machine Intelligence}}, vol.~21, no.~5, pp. 433--449, 1999.

\bibitem{Eleliemy:2017b}
A.~Eleliemy, A.~Mohammed, and F.~M. Ciorba, ``{Efficient Generation of Parallel
  Spin-images Using Dynamic Loop Scheduling},'' in \emph{{Proceedings of the
  19th IEEE International Conference for High Performance Computing and
  Communications Workshops (HPCCS 2017)}}, December 2017, p.~8.

\bibitem{SS}
T.~Peiyi and Y.~Pen-Chung, ``{Processor Self-Scheduling for Multiple-Nested
  Parallel Loops},'' in \emph{Proceedings of the International Conference on
  Parallel Processing}, August 1986, pp. 528--535.

\bibitem{FSC}
C.~P. Kruskal and A.~Weiss, ``{Allocating Independent Subtasks on Parallel
  Processors},'' \emph{{IEEE Transactions on Software Engineering}}, vol.
  SE-11, no.~10, pp. 1001--1016, 1985.

\bibitem{GSS}
C.~D. Polychronopoulos and D.~J. Kuck, ``{Guided Self-Scheduling: A Practical
  Scheduling Scheme for Parallel Supercomputers},'' \emph{{IEEE Transactions on
  Computers}}, vol. 100, no.~12, pp. 1425--1439, 1987.

\bibitem{FAC}
S.~Flynn~Hummel, E.~Schonberg, and L.~E. Flynn, ``Factoring: A method for
  scheduling parallel loops,'' \emph{Communications of the ACM}, vol.~35,
  no.~8, pp. 90--101, 1992.

\bibitem{SimGrid}
H.~Casanova, A.~Giersch, A.~Legrand, M.~Quinson, and F.~Suter, ``{Versatile,
  Scalable, and Accurate Simulation of Distributed Applications and
  Platforms},'' \emph{Journal of Parallel and Distributed Computing}, vol.~74,
  no.~10, pp. 2899--2917, 2014.

\bibitem{li1993locality}
H.~Li, S.~Tandri, M.~Stumm, and K.~C. Sevcik, ``{Locality and Loop Scheduling
  on NUMA Multiprocessors},'' in \emph{{Proceedings of the International
  Conference on Parallel Processing}}, August 1993, pp. 140--147.

\bibitem{surveyloop}
D.~F. Bacon, S.~L. Graham, and O.~J. Sharp, ``{Compiler Transformations for
  High-performance Computing},'' \emph{{ACM Computing Surveys}}, vol.~26,
  no.~4, pp. 345--420, 1994.

\bibitem{mahad}
M.~Balasubramanian, N.~Sukhija, F.~M. Ciorba, I.~Banicescu, and S.~Srivastava,
  ``{Towards the Scalability of Dynamic Loop Scheduling Techniques via Discrete
  Event Simulation},'' in \emph{{Proceedings of the International Parallel and
  Distributed Processing Symposium Workshops}}, May 2012, pp. 1343--1351.

\bibitem{WF}
S.~Flynn~Hummel, J.~Schmidt, R.~Uma, and J.~Wein, ``{Load-sharing in
  Heterogeneous Systems via Weighted Factoring},'' in \emph{{Proceedings of the
  Annual ACM Symposium on Parallel Algorithms and Architectures}}.\hskip 1em
  plus 0.5em minus 0.4em\relax ACM, June, 1996, pp. 318--328.

\bibitem{AF}
I.~Banicescu and Z.~Liu, ``{Adaptive Factoring: A Dynamic Scheduling Method
  Tuned to the Rate of Weight Changes},'' in \emph{{Proceedings of the High
  Performance Computing Symposium}}, April 2000, pp. 122--129.

\bibitem{AWFBC}
R.~L. Cari{\~n}o and I.~Banicescu, ``{Dynamic Load Balancing With Adaptive
  Factoring Methods in Scientific Applications},'' \emph{{Journal of
  Supercomputing}}, vol.~44, no.~1, pp. 41--63, 2008.

\bibitem{nitin}
N.~Sukhija, I.~Banicescu, S.~Srivastava, and F.~M. Ciorba, ``{Evaluating the
  Flexibility of Dynamic Loop Scheduling on Heterogeneous Systems in the
  Presence of Fluctuating Load Using {S}im{G}rid},'' in \emph{{Proceedings of
  the International Parallel and Distributed Processing Symposium Workshops}},
  May 2013, pp. 1429--1438.

\bibitem{dlsmsg}
N.~Sukhija, I.~Banicescu, and F.~M. Ciorba, ``{Investigating the Resilience of
  Dynamic Loop Scheduling in Heterogeneous Computing Systems},'' in
  \emph{{Proceedings of the International Symposium on Parallel and Distributed
  Computing}}, June 2015, pp. 194--203.

\bibitem{hoffeins2017examining}
F.~Hoffeins, F.~M. Ciorba, and I.~Banicescu, ``{Examining the Reproducibility
  of Using Dynamic Loop Scheduling Techniques in Scientific Applications},'' in
  \emph{{International Parallel and Distributed Processing Symposium
  Workshops}}, May 2017, pp. 1579--1587.

\bibitem{Ali_HPCC:17}
A.~Mohammed, A.~Eleliemy, and F.~M. Ciorba, ``{Towards the Reproduction of
  Selected Dynamic Loop Scheduling Experiments Using SimGrid-SimDag},'' Poster
  at IEEE International Conference on High Performance Computing and
  Communications (HPCC), 2017.

\bibitem{tesser2017}
R.~K. Tesser, L.~M. Schnorr, A.~Legrand, F.~Dupros, and P.~O.~A. Navaux,
  ``{Using Simulation to Evaluate and Tune the Performance of Dynamic Load
  Balancing of an Over-decomposed Geophysics Application},'' in
  \emph{{Proceedings of the International Conference on Parallel and
  Distributing Computing}}, August 2017, pp. 192--205.

\bibitem{AMPI}
C.~Huang, O.~Lawlor, and L.~V. Kale, ``{Adaptive MPI},'' in \emph{{Proceedings
  of the International Workshop on Languages and Compilers for Parallel
  Computing}}, October 2003, pp. 306--322.

\bibitem{TiT}
F.~Desprez, G.~S. Markomanolis, and F.~Suter, ``Improving the accuracy and
  efficiency of time-independent trace replay,'' in \emph{Proceedings of the
  International High Performance Computing, Networking, Storage and Analysis},
  November 2012, pp. 446--455.

\bibitem{SimGrid:Network}
P.~Bedaride, A.~Degomme, S.~Genaud, A.~Legrand, G.~S. Markomanolis, M.~Quinson,
  M.~Stillwell, F.~Suter, and B.~Videau, ``{Toward Better Simulation of MPI
  Applications on Ethernet/TCP Networks},'' in \emph{{Proceedings of the
  International Workshop on Performance Modeling, Benchmarking and Simulation
  of High Performance Computer Systems}}, November 2013, pp. 158--181.

\bibitem{huang2006performance}
C.~Huang, G.~Zheng, L.~Kal{\'e}, and S.~Kumar, ``{Performance Evaluation of
  Adaptive MPI},'' in \emph{Proceedings of the ACM SIGPLAN Symposium on
  Principles and Practice of Parallel Programming}, March 2006, pp. 12--21.

\bibitem{dataset}
K.~Wang, G.~Lavou\'e, F.~Denis, A.~Baskurt, and X.~He, ``{A benchmark for 3{D}
  mesh watermarking},'' in \emph{Proceedings of the 9th IEEE International
  Conference on Shape Modeling and Applications}, 2010, pp. 231--235.

\bibitem{le2010performance}
J.-Y. Le~Boudec, \emph{Performance evaluation of computer and communication
  systems}.\hskip 1em plus 0.5em minus 0.4em\relax EPFL Press, 2010.

\bibitem{Ali_SC:17}
A.~Mohammed, A.~Eleliemy, and F.~M. Ciorba, ``{A Methodology for Bridging the
  Native and Simulated Execution of Parallel Applications},'' Poster at
  ACM/IEEE International Conference for High Performance Computing, Networking,
  Storage, and Analysis, 2017.

\bibitem{papi}
S.~Browne, J.~Dongarra, N.~Garner, G.~Ho, and P.~Mucci, ``A portable
  programming interface for performance evaluation on modern processors,''
  \emph{{International Journal of High Performance Computing Applications}},
  vol.~14, no.~3, pp. 189--204, 2000.

\bibitem{simgrid_calib}
SimGrid, ``{SimGrid Calibration’s documentation},''
  http://simgrid.gforge.inria.fr/contrib/smpi-calibration-doc/, 2014, [Online;
  accessed 17 April 2018].

\bibitem{repro}
A.~Mohammed, A.~Eleliemy, F.~M. Ciorba, F.~Kasielke, and I.~Banicescu,
  ``{Experimental Verification and Analysis of Dynamic Loop Scheduling in
  Scientific Applications},''
  https://drive.switch.ch/index.php/s/xZD0gPPkTMWxZit, February 2018, [Online;
  accessed 26 February 2018].

\end{thebibliography}

\end{document}